# Colloidal dispersions of sterically and electrostatically stabilized PbS quantum dots: the effect of stabilization mechanism on structure factors, second virial coefficients, and film-forming properties

Ahhyun Jeong,[1,§] Josh Portner,[1,§] Christian P. N. Tanner,[2] Justin C. Ondry,[1] Chenkun Zhou,[1] Zehan Mi,[1] Youssef A. Tazoui,[1] Vivian R. K. Wall,[2] Naomi S. Ginsberg,[2-6] and Dmitri V. Talapin[1,7]*

[1]Department of Chemistry, James Franck Institute, and Pritzker School of Molecular Engineering, University of Chicago, Chicago, Illinois 60637, United States

[2]Department of Chemistry, University of California, Berkeley, California 94720, United States

[3]Department of Physics, University of California, Berkeley, California 94720, United States

[4]Molecular Biophysics and Integrated Bioimaging Division and Materials Sciences and Chemical Sciences Divisions, Lawrence Berkeley National Laboratory, Berkeley, California 94720, United States

[5]Kavli Energy NanoSciences Institute, University of California, Berkeley, California 94720, United State

[6]STROBE, NSF Science & Technology Center, Berkeley, California 94720, United States

[7]Center for Nanoscale Materials, Argonne National Laboratory, Argonne, Illinois 60439, United States

*E-mail: dvtalapin@uchicago.edu

[§]These authors contributed equally.

## Abstract

Electrostatically stabilized nanocrystals (NCs) and, in particular, quantum dots (QDs) hold promise for forming strongly coupled superlattices due to their compact and electronically conductive surface ligands. However, studies on the colloidal dispersion and interparticle interactions of electrostatically stabilized sub-10 nm NCs have been limited, hindering the optimization of colloidal stability and self-assembly. In this study, we employed small-angle X-ray scattering (SAXS) experiments to investigate the interparticle interactions and arrangement of PbS QDs with thiostannate ligands (PbS-$Sn_2S_6^{4-}$) in polar solvents. The study reveals significant deviations from ideal solution behavior in electrostatically stabilized QD dispersions. Our results demonstrate that PbS-$Sn_2S_6^{4-}$ QDs exhibit long-range interactions within the solvent, in contrast to the short-range steric repulsion characteristic of PbS QDs with oleate ligands (PbS-OA). Introducing highly charged multivalent electrolytes screens electrostatic interactions between charged QDs, reducing the length scale of the repulsive interactions. Furthermore, we calculate the second virial ($B_2$) coefficients from SAXS data, providing insights into how surface chemistry, solvent, and size influence pair potentials. Finally, we explore the influence of long-range interparticle interactions of PbS-$Sn_2S_6^{4-}$ QDs on the morphology of films produced by drying or spin-coating colloidal solutions. The long-range repulsive term of PbS-$Sn_2S_6^{4-}$ QDs promotes the formation of amorphous films, and screening the electrostatic repulsion by addition of an electrolyte enables the formation of a crystalline film. These findings highlight the critical role of NC-NC interactions in tailoring the properties of functional nanomaterials.

**Introduction**

Colloidal nanocrystals (NCs) are crystalline nanoscale particles that are dispersed in a solvent medium. The nanocrystals of direct-gap semiconductors, also known as quantum dots (QDs), are of high technological importance. For example, colloidal PbS QDs are widely explored for infrared photodetectors and solar cells;[1-3] PbS QDs are also studied for use in field-effect transistors,[4] light-emitting diodes,[3, 5] and thermoelectric devices.[4] A key advantage of QDs is the combination of physical properties inherent to inorganic semiconductors with inexpensive and scalable solution-based synthesis, processing and device integration methods more typically associated with molecular species and polymers.

In a colloidal solution, NCs stay separated from each other by repulsive pair potentials, which prevent unwanted NC aggregation. The repulsive pair potentials can be introduced by attaching organic ligands with long flexible chains (so-called "polymer brushes") to the NC surface, thus leading to steric repulsion between NCs.[6] An alternative approach involves replacing the bulky organic ligands with charged inorganic ligands such as halides (I⁻, Br⁻, Cl⁻) or metal chalcogenide complexes (e.g. $Sn_2S_6^{4-}$, $AsS_3^{3-}$, $GeS_4^{4-}$), which induce repulsion between nanoparticles through the electrostatic mechanism of colloidal stabilization.[7-10] In contrast to NCs sterically stabilized with bulky organic ligands, the use of compact inorganic ligands is appealing for optoelectronic applications because they enable NCs to pack more closely, thus enhancing the electronic coupling and increasing electron mobility by orders of magnitude.[11-13] This makes possible exciting new properties, such as electronic minibands in the ordered arrays of strongly electronically coupled NCs,[14, 15] as well as improved characteristics of optoelectronic devices.

A comprehensive understanding of the NC-NC interactions in colloidal solutions is essential for making nanocrystal devices by solution processing methods, such as spin-coating or inkjet printing, as the interparticle interactions exert a profound impact on colloidal stability and NC assembly. Previous studies have utilized scattering and microscopy methods to experimentally study the NC-NC interactions in sterically stabilized NCs, revealing the influence of size, surface, solvent conditions on the colloidal stability and assembly.[16-20] However, our understanding of NC-NC interactions in colloidal dispersions of electrostatically stabilized NCs remains limited. While the Derjaguin-Landau-Verwey-Overbeek (DLVO) theory offers a framework for modelling interparticle interactions of charged particles, it breaks down when the colloid is dispersed in a non-1:1 electrolyte solution or at high electrolyte concentrations.[21] Recent work by Coropceanu *et al.* has demonstrated that flocculating electrostatically stabilized NCs in a solution of high concentrations of multivalent electrolyte can induce self-assembly, which cannot be explained by the DLVO theory alone.[22-24] To develop functional materials that harness the unique properties of electrostatically stabilized NCs, a thorough understanding of how the NC size, surface chemistry and solvent influence the interactions between electrostatically stabilized NCs will be necessary to advance the field.

Here, we utilize small-angle X-ray scattering (SAXS) to study colloidal solutions of electrostatically stabilized PbS QDs. We show that sub-10 nm diameter PbS QDs colloidally stabilized with $Sn_2S_6^{4-}$ ligands (PbS-$Sn_2S_6^{4-}$) exhibit long-range interactions, resulting in complex interactions between the NCs within the solvent. In contrast, sterically stabilized PbS QDs with oleate (OA) ligands (PbS-OA) only exhibit short-range repulsions not extending beyond the full ligand length, and hence such colloids act as hard spheres across a broad range of QD concentrations. We show that with the addition of a highly charged multivalent electrolyte, we can screen the electrostatic interactions between charged nanocrystals and recover a hard-sphere behavior for electrostatically stabilized QD colloids. Further analysis of SAXS data enabled us to determine the second virial ($B_2$) coefficients to quantitatively examine how surface chemistry, solvent, and composition of QDs influence their repulsive pair potentials. From these insights, we are able to identify a strategy to prepare films of strongly coupled and highly ordered superlattices of PbS QDs capped with inorganic metal chalcogenide complex surface ligands.

**Results and Discussion**

**Preparation and surface characterization of PbS QDs.** PbS QDs with precisely controlled size and narrow size distributions are synthesized from lead oleate and substituted thioureas according to the methods developed by the Owen group.[25] The nanocrystals used in this study have diameters of 4.0 to 8.3 nm (See SI Section 2 and Figure 1A). PbS QDs with thiostannate ligands are prepared by performing a bi-phasic ligand exchange on PbS-OA QDs using potassium thiostannate as a ligand source and *N*-methylformamide as a solvent. With this ligand exchange procedure, we produce two nanocrystal populations with identical QD cores but distinct surface chemistries (Figure 1B). The PbS-OA samples are stabilized by steric repulsion between long-chain organic ligands and the thiostannate-capped PbS samples are stabilized by the electrostatic mechanism.

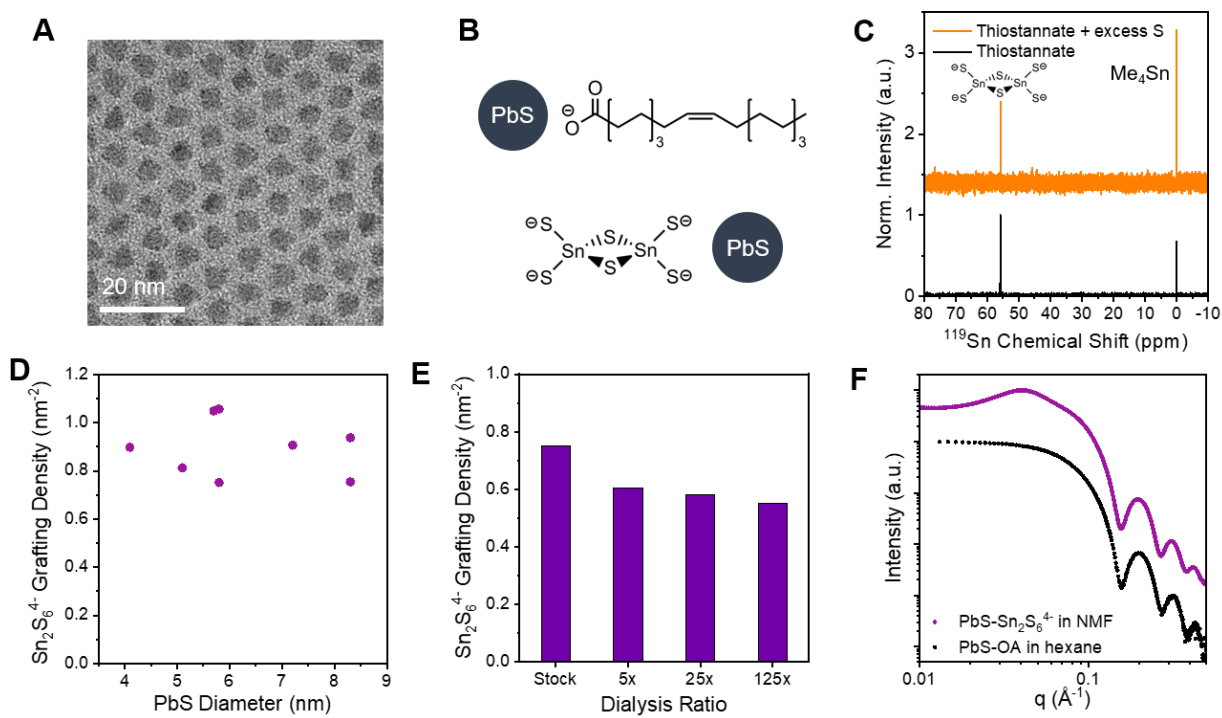

**Figure 1. (A)** TEM images of 5.7 nm PbS-OA QDs. **(B)** The chemical structures of oleate (OA) and $Sn_2S_6^{4-}$ ligands on a surface of PbS QDs. **(C)** $^{119}$Sn NMR of thiostannate salts with and without excess sulfide in NMF. **(D)** Grafting density of $Sn_2S_6^{4-}$ ligands on PbS NCs of different sizes. **(E)** Grafting density of $Sn_2S_6^{4-}$ ligands on 5.8 nm PbS NCs following a series of dialysis cycles. **(F)** Representative log-log plot of SAXS patterns of 5.7 nm PbS-OA and PbS-$Sn_2S_6^{4-}$ QDs. The patterns are shifted vertically for clarity of presentation.

Various thiostannate species can exist in polar solvents ($Sn_2S_6^{4-}$, $Sn_2S_7^{6-}$, $SnS_4^{4-}$), so it is necessary to confirm the identity of the thiostannate ligand coordinated to PbS QDs upon ligand exchange. To verify the identity of the thiostannate species present in NMF, we conducted $^{119}$Sn nuclear magnetic resonance (NMR) studies. When the thiostannate salt was prepared by dissolving $K_2S$ and $SnS_2$ in NMF at 2:1 molar ratio, only one peak at 55 ppm was observed, corresponding to the $K_4Sn_2S_6$ species. No other species were detected in the NMR spectrum (e.g. $K_6Sn_2S_7$ at 68 ppm, $K_4SnS_4$ at 74 ppm).[26] Even after adding excess $K_2S$ to increase the sulfur ratio, no alternative species were formed. Consequently, we concluded that $Sn_2S_6^{4-}$ is the predominant species serving as the ligand for thiostannate-capped PbS QDs in NMF, counter-balanced by the $K^+$ ions in the solution.

Figure 1D displays the grafting density of $Sn_2S_6^{4-}$ ligands on the surface of 5.8 nm PbS QDs, determined through inductively coupled plasma optical emission spectroscopy (ICP-OES) measurement. As shown in Figure 1E, elemental analysis of PbS NCs with different sizes reveals that the grafting density of $Sn_2S_6^{4-}$ on PbS NCs before the dialysis varies from 0.75 to 1.05 ligands per nm$^2$ and that there is no clear correlation between the size of PbS NCs and the $Sn_2S_6^{4-}$ ligand grafting density. To determine the concentrations of free and bound ligands, a series of dialysis experiments using centrifugal dialysis filters were conducted. The dialysis filters allow solvent and free ligands to diffuse through the membrane pores, while the larger NCs remain trapped. As an example, the 5x dialysis cycle was performed by removing 4/5 of the solvent and free ligands using a dialysis filter, and then adding the equal volume of pure NMF to restore the original concentration of NCs while reducing the free ligand concentration by 5 times. A dialysis ratio of 125x was achieved by repeating the 5x dialysis cycle three times. The results indicate that the grafting density of $Sn_2S_6^{4-}$ ligands on 5.8 nm PbS QDs decreases from 0.75 to 0.55 per nm$^2$ after 125x dialysis, suggesting that approximately 0.55 per nm$^2$ of $Sn_2S_6^{4-}$ ligands are strongly bound to the surface out of the initial grafting density of 0.75 per nm$^2$.

Figure 1F shows representative SAXS patterns of sterically (PbS-OA) and electrostatically (PbS-$Sn_2S_6^{4-}$) stabilized PbS QDs. The scattering intensity *vs.* momentum transfer ($q$) dependence in SAXS patterns of PbS-OA exhibits distinct Bessel oscillations at large $q > 0.1$ Å$^{-1}$, indicative of narrow size and shape distributions of the QDs.[27] This observation remains consistent after the ligand exchange process, as evidenced by sharp Bessel oscillations in the SAXS pattern of PbS-$Sn_2S_6^{4-}$ samples. In addition, both PbS-OA and PbS-$Sn_2S_6^{4-}$ QDs do not exhibit an up-turn at low-$q$ region (q < 0.1 Å$^{-1}$), suggesting that both types of QDs are well-dispersed and colloidally stable in solution.[27] Unstable solutions of QDs exhibit an up-turn at low-$q$ region, as illustrated in Figure S8. However, the low-$q$ regions of the SAXS patterns exhibit notable differences between the two types of QDs. At small $q$-values, the scattering intensity of PbS-OA QDs converges to a horizontal line on a log-log scale (Guinier region), while the scattering intensity of PbS-$Sn_2S_6^{4-}$ QDs clearly deviates from a horizontal line. This indicates that colloidal dispersions of sterically and electrostatically stabilized PbS QDs have distinct structures. To examine the unusual features of the SAXS patterns of PbS-$Sn_2S_6^{4-}$ QDs in the low $q$ region, we extract the structure factor. The SAXS intensity $I(q) = A\, P(q)\, S(q)$, where $A$ is the scaling factor, $P(q)$ is the form factor, and $S(q)$ is the structure factor. The scaling factor is dependent on the measurement parameters, such as the intensity of incident X-rays, concentration and chemical composition of NCs.[27] The form factor is defined by the geometric properties of NCs, such as their size, shape and polydispersity. $P(q)$ can be estimated by fitting the Porod region ($qR \gg 1$), where $R$ is NC radius, with the equation of form factor for spherical particles (SI Section 3.1).[27]

**Structure factors of colloidal solutions.** The structure factor $S(q)$ is dependent on the spatial arrangement of NCs. It provides valuable insights into the interactions between the NCs in a colloidal dispersion. If interparticle interactions are negligible, the NCs adopt a random arrangement, resulting in a structure factor close to unity for different $q$-values. Conversely, when strong interparticle interactions are present, the NCs deviate from ideal gas behavior, leading to distinct features in the structure factor.[28] By analyzing $S(q)$, we can gain a better understanding of internal structure of the NC colloid and relate it to the interparticle interactions.

Figure 2A presents a comparison of the structure factors for PbS-OA and PbS-Sn$_2$S$_6^{4-}$ QDs. The intensity of oscillation in the structure factor of PbS-OA QDs is observed to be smaller compared to that of PbS-Sn$_2$S$_6^{4-}$ QDs, indicating a more randomized distribution of particles within the colloidal solution (Figure 2E). This implies that the particles experience interparticle forces that are weak or effective only over short distances, effectively behaving like hard spheres (SI Section 3.3). Such behavior is expected for PbS-OA QDs, which interact only through the van der Waals attraction and steric repulsion.[29] This behavior is in agreement with that of other sterically stabilized colloid systems.[30] On the other hand, the structure factor of electrostatically stabilized PbS QDs present pronounced oscillatory features, indicating deviation from typical gas-like behavior. For PbS-Sn$_2$S$_6^{4-}$ QDs in Figure 2A, the structure factor shows two key characteristics: i) $S(q)$ drops well below unity at low q, indicating that the QD colloids lack density fluctuations over large length scales, and ii) a peak at 0.045 Å$^{-1}$ suggests that the particles are uniformly separated by the nearest neighboring center-to-center distance ($d$) of ~14 nm, as calculated by $d \approx 2\pi/q_{max}$.[31] This implies that electrostatically stabilized NCs experience long-ranged repulsive forces acting on neighboring particles, resulting in the uniform arrangement illustrated in Figure 2E.[28, 32]

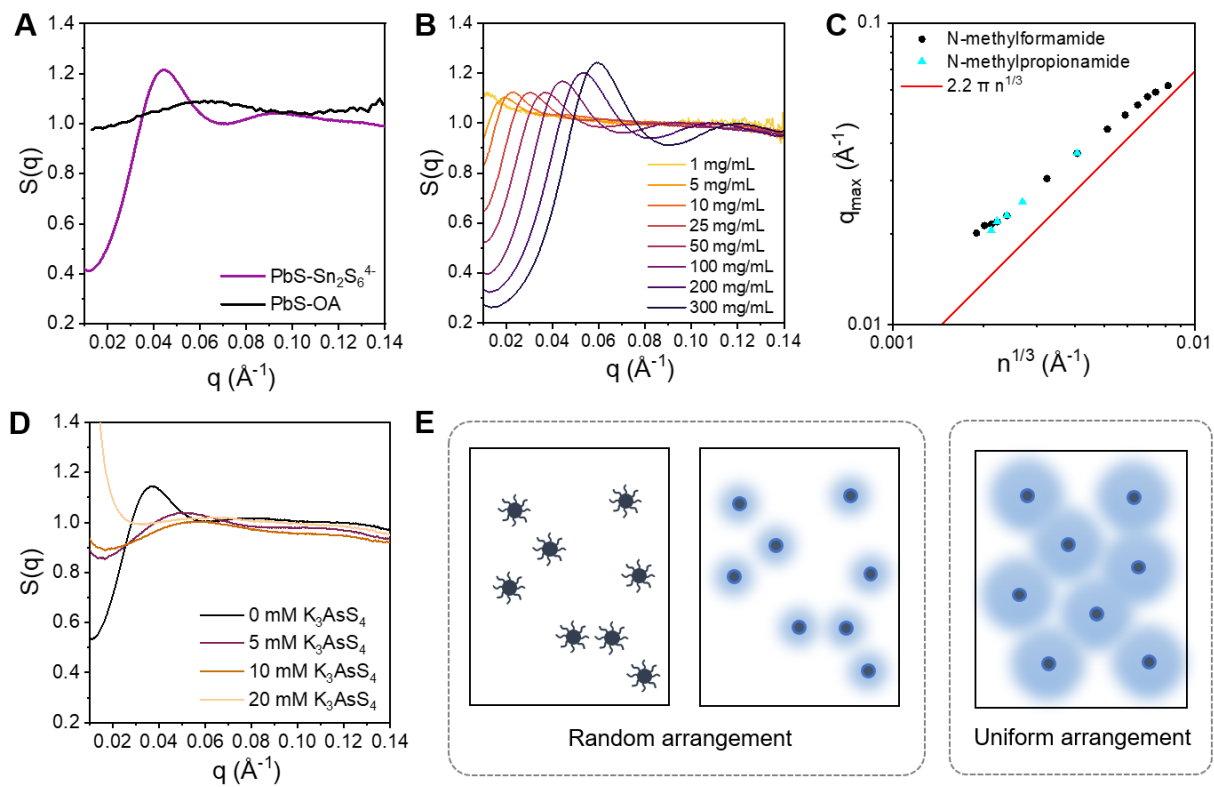

**Figure 2. (A)** The structure factors of colloidal dispersions of 50 mg/ml of 5.7 nm PbS QDs with different surface ligands: PbS-$Sn_2S_6^{4-}$ QDs in NMF (blue) and PbS-OA in hexane (red). **(B)** The structure factor of 5.7 nm PbS-$Sn_2S_6^{4-}$ QDs in NMF at different concentrations. **(C)** The plot of $q_{max}$ of 5.7 nm PbS-$Sn_2S_6^{4-}$ QDs against $n^{1/3}$ in NMF and NMPA solvents. The red line ($q_{max} = 2.2\pi n^{1/3}$) indicates the theoretical trend when NCs are separated by greatest possible distance. **(D)** The structure factors of 50 mg/ml of 5.7 nm PbS-$Sn_2S_6^{4-}$ QDs in NMF at different concentrations of added $K_3AsS_4$ salt. **(E)** Illustration of random arrangements of PbS-OA QDs in hexane (left), PbS-$Sn_2S_6^{4-}$ QDs in NMF with added $K_3AsS_4$ salts (middle), and the illustration of a uniform arrangement of PbS-$Sn_2S_6^{4-}$ QDs at low salt concentration (right).

Figure 2B demonstrates the pronounced impact of the concentration of electrostatically stabilized NCs on the structure factor. As the QD concentration rises from 1 mg/ml to 300 mg/ml, the first maximum of $S(q)$ shifts from $q = 0.012$ Å$^{-1}$ to $q = 0.060$ Å$^{-1}$, signifying a reduction in the characteristic center-to-center separation between adjacent NCs from 52 nm to 10 nm (Figure S11). In Figure 2C, the correlation between the peak position of $S(q)$ and $n^{1/3}$ is depicted, where $n$ is the number density of the QDs in colloidal dispersions, and $n^{-1/3}$ represents the mean center-to-center separation of particles.[33] Computational studies have shown that the peak position of the structure factor ($q_{max}$) of the suspension of strongly repulsive particles with Yukawa potential scales according to the relationship $q_{max} \approx 2.2\pi n^{1/3}$.[33] A strong correlation between the $q_{max}$ of PbS-$Sn_2S_6^{4-}$ QDs in NMF and *N*-methylpropionamide (NMPA) and $n^{1/3}$ is observed, although the trend does not precisely align with $q_{max} = 2.2\pi n^{1/3}$ (see Figure S12 for the structure factors). The deviation originates from the increased concentration of counter-ions ($K^+$) in a concentrated solution of QDs. In a concentrated solution of QDs, the electrolytes in solution screen the surface potential of the QDs, reducing the Debye length and allowing the QDs to approach each other more closely than $n^{-1/3}$.[33]

In Figure 2B, we also observe that the peak-to-valley ratio of the structure factor rises with increasing concentration. The ratio of the value of structure factor at $q = 0$, i.e., $S(0)$ and the peak value of structure factor ($S_{max}$) is known as the hyperuniformity index, which quantifies the spatial uniformity of the arrangement of NCs.[34] At low concentrations, NC-NC repulsions are weak due to an exponential decay of electrostatic repulsion with distance. This allows the NCs to adopt random arrangement, reflected by the near-unity value of $S(0)/S_{max}$. On the other hand, in a concentrated solution of PbS QDs, strong electrostatic repulsions force the NCs to be separated by relatively uniform distances, as evidenced by the intense primary peak and the depression of $S(q)$ at small $q$ values (Figure S13). Despite this uniform arrangement, the QDs remain far from a "jammed" structure, where QD movement is restricted by repulsive forces from neighboring particles. X-ray photon correlation spectroscopy (XPCS) measurements confirm that the diffusion rate of QDs in a dilute and concentrated solutions of PbS-$Sn_2S_6^{4-}$ QDs show negligible differences (Figure S14).[35] This is likely due to the insufficient strength of interparticle repulsion of the QDs to confine the movement of QDs at these volume fractions. The uniformity of QDs separated by soft and long-range repulsive potentials holds promise for producing disordered but highly uniform materials, useful for application in fields such as optically isotropic waveguides and sources of isotropic thermal radiation.[34, 36]

The addition of free ions to an electrostatically stabilized PbS-$Sn_2S_6^{4-}$ solution is expected to reduce the screening length of electrostatic repulsion.[37] Consequently, the long-range repulsion of NCs is alleviated, resulting in the reappearance of a colloidal gas-like configuration (Figure 2E). This phenomenon is corroborated by the structure factors shown in Figure 2D, where the oscillatory features of structure factor are suppressed by the addition of K₃AsS₄ salt. The addition of 10 mM K₃AsS₄ results in near-unity $S(q)$, similar to sterically stabilized PbS-OA QDs. Notably, the rise of structure factors in the low-$q$ region may be attributed to partial aggregation of PbS QDs, forming transient aggregates that yield strong X-ray scattering in this range.[31] An alternative explanation could involve the creation of clusters of NCs driven by the attraction of negatively charged NCs to positively charged counterions encircling the neighboring NCs.[38] The cluster fluid, percolated fluid or periodic microphase configurations of NCs, predicted for colloids of particles with short-range attractive and long-range repulsive (SALR) interactions, could give rise to an additional peak in the low-$q$ region of the structure factor as well.[39]

**Second virial coefficients.** To quantitatively analyze how the size, surface, and solvent of NCs impact the extent of deviation from the ideal behavior, we measured the second virial ($B_2$) coefficients of colloidal dispersions of electrostatically stabilized PbS QDs. Various physical properties of non-ideal solution can be defined through virial coefficients. Thus, the osmotic pressure ($\Pi$) of a dispersion of interacting particles can be expressed as a power series of particle number density ($n$), where $k_B$ is the Boltzmann constant, $T$ is the temperature, $B_2$ is the second virial coefficient and $B_3$ is the third virial coefficient:[40]

$$\frac{\Pi}{k_B T} = n + B_2 n^2 + B_3 n^3 + \ldots \qquad (1)$$

A positive $B_2$ coefficient indicates a net repulsive, and a negative $B_2$ coefficient indicates a net attractive pairwise interaction between the particles. Quantitatively, $B_2$ can be directly related to the pair potentials $u(r)$ between NCs that are separated by the distance $r$:[30]

$$B_2 = -2\pi \int_0^\infty \left(\exp\left(-\frac{u(r)}{k_B T}\right) - 1\right) r^2 \, dr \qquad (2)$$

When comparing $B_2$ coefficients of different NC samples, it is often convenient to calculate the normalized second virial coefficients ($b_2$), which provide more direct insights into colloidal stability and self-assembly behavior. For spherical or nearly spherical particles, a dimensionless $b_2$ can be calculated as $b_2 = B_2/B_{2,HS}$, where $B_{2,HS}$ is the second virial coefficient of an equivalent colloid of hard spheres. In turn, $B_{2,HS}$ can be calculated as $B_{2,HS} = 2\pi(d + 2\delta)^3/3$, where $d$ is the diameter of the NC core and $\delta$ is the length of the surface ligands.[30] A positive $b_2$ value implies an overall repulsive interactions between NCs, leading to high colloidal stability. In contrast, a negative $b_2$, especially $b_2 < -10$, typically leads to uncontrolled aggregation into an amorphous structure.[41] Weakly attractive particles with a small $b_2$ coefficient ($-10 < b_2 < -1$) can self-assemble into ordered structures, such as protein crystals or NC superlattices.[41]

We experimentally determine the $B_2$ coefficients for PbS QDs from concentration-dependent SAXS data. The relationship between the number density ($n$) of NCs and the $B_2$ coefficient is described by equation 3, where $S(0)$ is the extrapolated value of structure factor at $q = 0$ and $\mathcal{O}(n^2)$ collects the higher-order terms in $n$:[30]

$$\frac{1}{S(0)} = 1 + 2B_2 n + \mathcal{O}(n^2) \tag{3}$$

The value of $S(0)$ can be obtained by performing a linear fit of $\ln[S(0)]$ against $q^2$ at small $q$ values (Figure S15).[42] Figure 3A displays representative plots of $S(q)$ and the corresponding fits for $S(0)$ in the low-$q$ regions for a series of concentrations of 7.2 nm PbS QDs with $Sn_2S_6^{4-}$ ligands in NMF. Figure 3B shows a representative plot of $1/S(0) - 1$ against the NC number density ($n$), allowing the $B_2$ coefficient to be determined from the slope. The structure factor and $B_2$ analysis of all other QDs are shown in Figure S16.

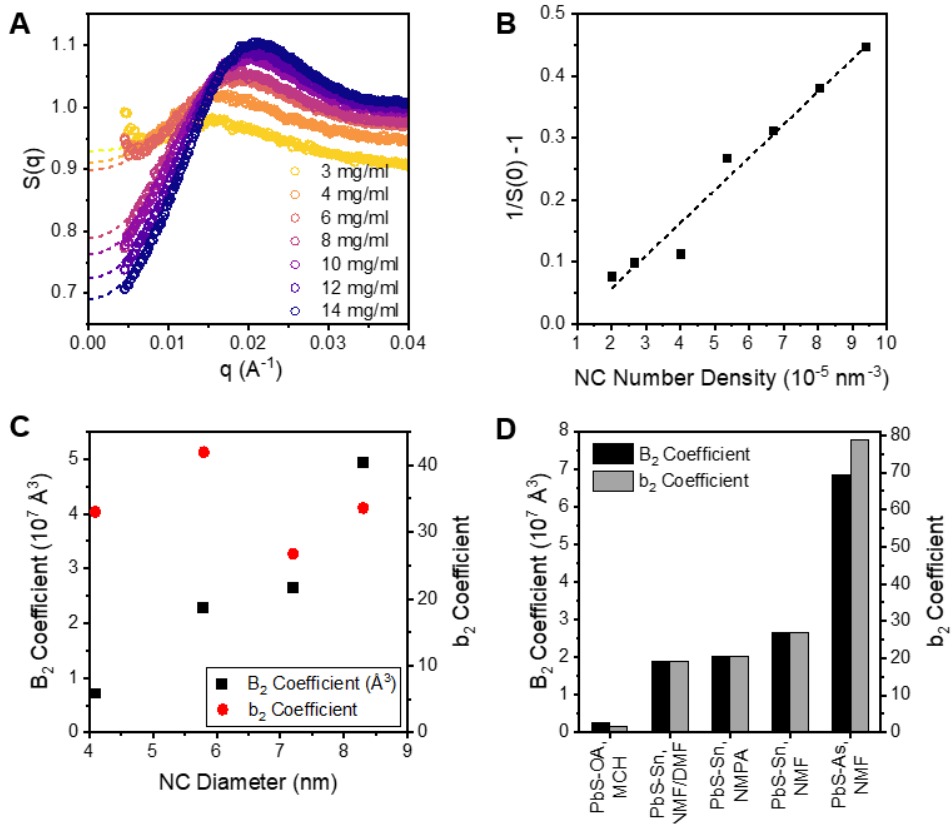

**Figure 3. (A)** $S(q)$ of 7.2 nm PbS QDs with $Sn_2S_6^{4-}$ ligands in NMF. The dashed lines are fits of $S(q)$ with $S(q) = A\exp(kq^2)$ in low-q region. The circles are experimentally determined $S(q)$ values. **(B)** The corresponding plot of $1/S(0) - 1$ against the number density of PbS QDs. **(C)** $B_2$ and $b_2$ coefficient of PbS-$Sn_2S_6^{4-}$ QDs in NMF with size ranging from 4.1 nm to 8.3 nm. **(D)** The $B_2$ and $b_2$ coefficients of ~7 nm PbS QDs under different surface chemistry and solvent conditions. The diameter of PbS-OA NCs is 6.9 nm, and the diameters of PbS-$Sn_2S_6^{4-}$ and PbS-$AsS_4^{3-}$ NCs are 7.2 nm. Here, Sn indicates $Sn_2S_6^{4-}$ ligands, As indicates $AsS_4^{3-}$ ligands and OA indicates oleate ligands. NMF/DMF is a mixture of NMF and DMF in a 7:3 volume ratio.

Figure 3C presents the $B_2$ and $b_2$ coefficients of four different PbS-$Sn_2S_6^{4-}$ QDs in NMF, ranging in size from 4.1 nm to 8.3 nm. Notably, the $B_2$ coefficient shows a monotonic increase with QD diameter. This can be attributed to the expanding volume occupied by each QD as its size increases, leading to increased steric repulsion. Additionally, the larger QDs have a greater surface area covered by negatively charged ligands, resulting in increased overall electrostatic repulsion in the surrounding space. This trend is consistent with the predictions from DLVO potential (Figure

S17), although the experimental values of $B_2$ coefficients exhibit stronger dependence on the size of the NCs compared to the predicted values from DLVO theory. Conversely, the $b_2$ coefficient shows only weak dependence on the QD diameter, indicating that the $B_2$ coefficient scales linearly with the volume of the QD.

In figure 3D, we compare the $B_2$ and $b_2$ coefficients of ~7 nm PbS QDs under different surface chemistry and solvent conditions. For the comparison, we focus on the $b_2$ coefficient, which shows little correlation with the size of PbS-$Sn_2S_6^{4-}$ QDs within the range of 4.1 – 8.3 nm (Figure 3C). Notably, all QDs exhibit positive $b_2$ coefficients, indicating the presence of repulsive interactions among the particles. The 7.2 nm PbS QDs that are electrostatically stabilized by $Sn_2S_6^{4-}$ and $AsS_4^{3-}$ ligands ($b_2 > 19$) demonstrate significantly larger $b_2$ coefficients compared to the sterically stabilized 6.9 nm PbS-OA NCs ($b_2 \approx 1.7$). This observation is consistent with the finding that the electrostatically stabilized NCs deviate more from the ideal gas behavior compared to the sterically stabilized NCs due to their long-range interactions. Additionally, the results reveal that the interparticle repulsion of PbS-$AsS_4^{3-}$ NCs is notably stronger than PbS-$Sn_2S_6^{4-}$ NCs. This can be attributed to the higher grafting density of $AsS_4^{3-}$ ligands on the NCs' surface (~2.25 nm$^{-2}$) compared to $Sn_2S_6^{4-}$ ligands (~0.55 nm$^{-2}$), resulting in the greater negative surface charge (Figure 1F and SI Section 6). Alternatively, the differences may arise from variations in the ionic strength of the solution due to unbound ligands present. The $Sn_2S_6^{4-}$ salts have higher valency than $AsS_4^{3-}$, screening the PbS-$Sn_2S_6^{4-}$ NCs more than the PbS-$AsS_4^{3-}$ NCs, resulting in weaker electrostatic repulsions between the NCs (Figure S18).

The $b_2$ coefficients of PbS-$Sn_2S_6^{4-}$ NCs were measured in NMF, *N*-methylpropionamide (NMPA) and a mixture of NMF and *N,N*-dimethylformamide (DMF) in a 7:3 volume ratio (Figure 3D). The observed trend of the $b_2$ coefficients is as follows: NMF > NMPA > NMF/DMF mixture. This trend follows the same order as the dielectric constants of the solvents, where NMF has the highest dielectric constant ($\varepsilon_{NMF} \sim 170$), followed by the NMPA ($\varepsilon_{NMPA} \sim 164$) and then the NMF/DMF mixture ($\varepsilon_{NMF/DMF} \sim 130$).[43, 44] This trend aligns with our expectation, as stronger ion-dipole interactions form between the surface ligands of the PbS-$Sn_2S_6^{4-}$ NCs and polar solvent molecules (Figure S19), as shown by the predicted trends from the DLVO theory. These interactions contribute to the enhanced colloidal stability of the NCs in solvents with high dielectric constant.

**Films formed from colloids of sterically and electrostatically stabilized NCs.** In the previous sections we showed that colloidal dispersions of electrostatically stabilized nanocrystals exhibit long-range repulsive forces. Within the framework of DLVO theory, the range of the repulsive forces is defined by the Debye screening length, and addition of an electrolyte into colloidal dispersion increases ionic strength and reduces Debye length.[21] According to several theoretical and computational studies,[45-50] the long-ranged electrostatic interactions between NCs should impact rheological properties and many other characteristics of colloidal dispersions.

The exact form of pair potentials for colloidal sub-10 nm NCs cannot be easily extracted from available experimental data.[51] It is generally accepted that pair potentials of electrostatically stabilized NCs should include at least three components – electrostatic repulsion, van der Waals (vdW) attraction, and a steep short-range repulsion at hard contact of NC cores. Although exact parameters of NC pair potentials may be difficult to access, it has been demonstrated that the behavior of $PbS-Sn_2S_6^{4-}$ NCs at high ionic strengths resembles the behavior of spheres with a pair potential consisting of a narrow attractive well with the width less than 20% of NC diameter and depth of several $k_BT$.[22-24] At low ionic strength, the interparticle interactions can be modeled as a combination of short-range attractive and longer-range repulsive (SALR) potentials, with the possibility of changing the contributions of attractive and repulsive components by varying the concentration of free ions added to a colloidal dispersion (Figure 4A). In contrast, sterically stabilized NCs in non-polar solvents behave very nearly as hard spheres.[30] Only when the solvent evaporates do the hydrocarbon chains of ligands begin experiencing strong van der Waals attractions.[52]

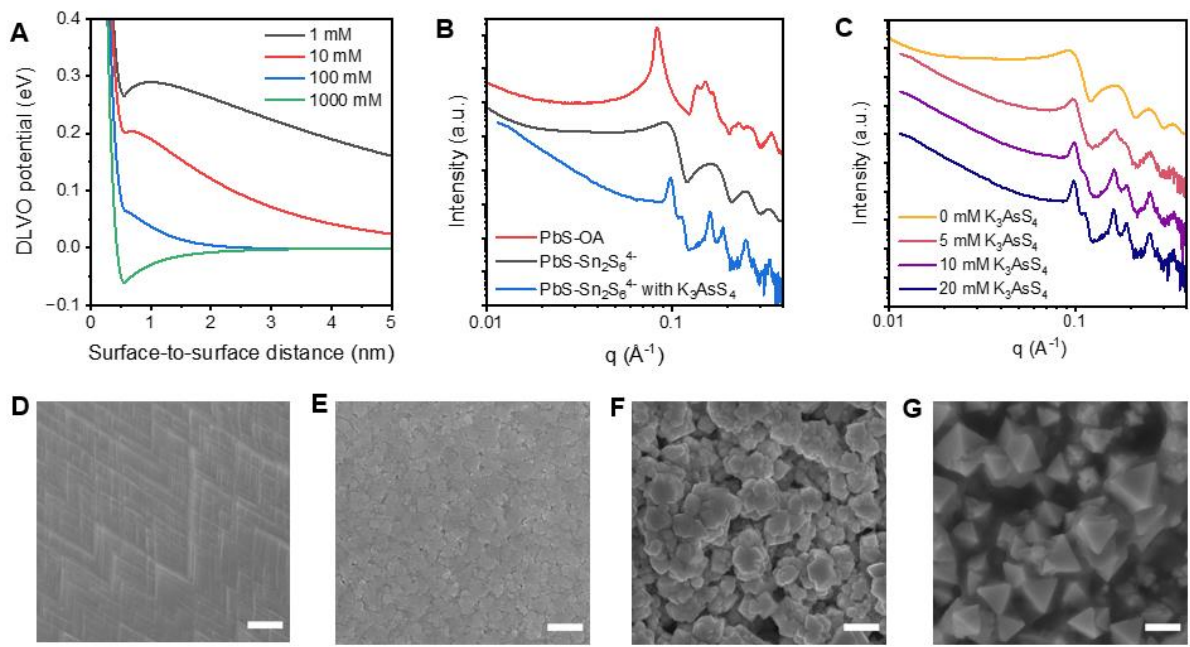

**Figure 4. (A)** DLVO potentials of 7.2 nm PbS-$Sn_2S_6^{4-}$ QDs in NMF with varying concentrations of added 1:3 electrolyte, e.g., $K_3AsS_4$ salt. **(B)** SAXS patterns of films of 7.2 nm PbS QDs with different ligands drop-casted on a thin Si wafer. **(C)** SAXS patterns of PbS-$Sn_2S_6^{4-}$ QD films drop-cast on a thin Si safer, with a varying concentrations of $K_3AsS_4$ added. **(D-G)** SEM images of films of **(D)** PbS-OA QDs, **(E)** PbS-$Sn_2S_6^{4-}$ QDs, **(F)** PbS-$Sn_2S_6^{4-}$ QDs with added 20 mM $K_3AsS_4$, **(G)** PbS-$Sn_2S_6^{4-}$ QDs with added 86 mM $K_3AsS_4$. All films were prepared by drop-casting. Scale bars in SEM images are 1 μm.

Given the importance of colloidal QDs for thin-film optoelectronic devices, we aim to compare the films of PbS QDs prepared from colloidal solutions of sterically and electrostatically stabilized colloidal dispersions and evaluate what characteristics of the pair potentials are most favorable for the formation of dense and uniform layers desirable for LED's, photodetectors, solar cells, and other thin-film devices.

Drop-casting or spin-coating of colloidal solutions of sterically stabilized PbS-OA QDs typically results in ordered films containing multiple superlattice domains. The domain size is defined by solvent evaporation rate and is smaller for spin-coated films compared to that in films made by slow solvent evaporation (Figures 4D and S23). Interestingly, the phase of superlattices can be either face-centered cubic (*fcc*) or body-centered cubic (*bcc*). For example, our films of 7.2 nm PbS-OA QDs deposited from methylcyclohexane have *fcc* structure (Figure 4B and S24). In several theoretical studies which related pair potentials and phase diagrams of colloidal spheres, it has been shown that *fcc* phase is more favorable for particles that interact by hard sphere-like potential, due to the higher packing density. Thus, very similar 7 nm PbS-OA QDs in toluene formed *fcc* superlattices upon slow addition of ethanol, which increased solvent polarity and gradually changed repulsive pair potentials to attractive.[53]

In contrast, electrostatically stabilized PbS-$Sn_2S_6^{4-}$ QDs show more complex behavior. In the absence of added ions, spin-coated or drop-cast colloidal dispersions of PbS-$Sn_2S_6^{4-}$ QDs in NMF yield glassy films with no QD ordering (Figures 4E and S24).[54] The formation of disordered films is consistent with theoretical predictions for particles interacting through SALR potentials.[45] In such systems, when the NC concentration increases due to solvent evaporation, the narrow attractive well promotes the formation of NC clusters which cannot grow further into superlattices due to the repulsive term.[46, 48] Upon further increase of NC concentration, these clusters coalesce forming percolated networks of NCs and even microphases.[39] The computational studies also predict that NC colloids with long-ranged SALR potentials are prone to gelation, also leading to disordered NC layers.[48] The randomly packed glassy NC layers are expected to have more isotropic properties compared to crystalline superlattices, which can be useful for applications in optoelectronic devices.

The addition of free ions, e.g., by adding $K_3AsS_4$ salt to the colloidal dispersion, whose concentration is gradually increasing upon solvent evaporation, is expected to suppress both the height and the range of repulsive component as shown in Figure 4A. As it was shown in our previous study, elimination of the repulsive term is favorable for self-assembly of NCs into long-range ordered *fcc* superlattices[22] because nucleated clusters of NCs can keep growing indefinitely. Figures 4C and S24 show that the addition of $K_3AsS_4$ salt to 7.2 nm PbS-$Sn_2S_6^{4-}$ QDs increases ordering of QDs in both drop-casted and spin-coated films. Ultimately, the film becomes a layer of faceted superlattices of PbS QDs, as shown in Figure 4F and 4G. Individual superlattices have nearly perfect octahedral faceting, corresponding to the Wulff construction with low energy facets with $(111)_{SL}$ Miller indices. The formation of such structures is expected if NCs have strong tendency to self-assemble into crystalline superlattices, while maximization of total cohesive energy through densification of NC layer is not happening. In one plausible scenario, superlattices of PbS-$Sn_2S_6^{4-}$ QDs nucleate and grow at the early stage of film drying, and the precipitation of micron-size superlattices results in a pile of facetted superlattices (Figure 4G and S25). Naturally, the long-range electronic connectivity for such film morphology is relatively weak, and, despite the possibility for interesting physics (e.g. miniband formation) within superlattices of strongly electronically coupled QDs,[14] the utility for such layers for optoelectronic applications is questionable.

Varying the concentration and nature of free ions in solution allows tuning of the pair potentials from long-ranged strongly repulsive, resulting in amorphous NC films, to short-range attractive that yields perfect but loosely packed QD superlattices. In between of these limiting cases, these can be a "sweet spot" where QDs self-assemble into ordered domains that are shaped to connect with neighboring QD superlattices without gaps. Such a morphology maximizes packing density and electronic coupling both within and between superlattice domains and enables good electronic connectivity across the entire NC film. Indeed, using a small concentration of free ions (no added $K_3AsS_4$ salt, but also no excessive washing of 7.2 nm PbS-$Sn_2S_6^{4-}$ QDs, which removes trace residual electrolyte) allowed us to achieve the morphology shown in Figure S25. It demonstrates that electrostatically stabilized colloidal dispersions of semiconductor QDs can be used as "inks" for making QD layers with packing density and structural quality comparable to those in films prepared from colloids of sterically stabilized NCs.

**Conclusion.** In this study, we compare the interparticle interactions of sterically and electrostatically stabilized PbS QDs using SAXS measurements and their influence on the structure and film-forming properties of the PbS QDs. The comprehensive study of the structure factors has revealed that the electrostatically stabilized QDs exhibit strong QD-QD interactions that persist over long distances, a characteristic absent from the conventional sterically stabilized QDs. However, the introduction of salts like $K_3AsS_4$ can reduce this interaction range, restoring short-range interactions similar to those in sterically stabilized QDs. Further analysis of the structure factors of electrostatically stabilized colloidal dispersions of PbS QDs with different sizes (4.1 – 8.3 nm) and surface chemistries ($Sn_2S_6^{4-}$, $AsS_4^{3-}$) and in different polar solvents (NMF, NMPA, DMF) revealed that the solutions of electrostatically stabilized QDs exhibit significant deviation from ideal solution behavior. The correlation between the experimental $B_2$ coefficients with various factors were found to be qualitatively consistent with the DLVO model, although the experimental $B_2$ coefficients were consistently larger than the values predicted from DLVO model. This may be attributed to the complex interactions of counter-ions ($K^+$) and multivalent ions ($Sn_2S_6^{4-}$), which cannot be fully captured by the DLVO model.

Our results provide insight into the influence of interparticle interactions in the morphology of thin films. Weakly repulsive particles, such as sterically stabilized PbS-OA QDs, tend to form crystalline or polycrystalline films, while the strongly repulsive particles like electrostatically stabilized PbS-Sn$_2$S$_6$ QDs tend to form amorphous films. The addition of K$_3$AsS$_4$ salt reduces the interactions and enables crystalline film formation for PbS-Sn$_2$S$_6$ QDs. These findings underscore the importance of QD-QD interactions in tailored film deposition and functional nanomaterials.

## ACKNOWLEDGEMENTS

We are grateful to Dr. Andrew Nelson for a critical reading and editing of the manuscript. We appreciate Dr. Byeongdu Lee and Dr. Jack Douglas for fruitful discussions. We acknowledge Dr. Xiaobing Zuo and Dr. Byeongdu Lee for their support for X-ray scattering experiments in Advanced Photon Source, Argonne National Laboratory. We acknowledge the European XFEL in Schenefeld, Germany, for provision of X-ray free electron laser beamtime at Scientific Instrument MID (Materials Imaging and Dynamics) and would like to thank the staff for their assistance. Material synthesis, X-ray scattering, X-FEL experiments, and simulations were supported by the Office of Basic Energy Sciences (BES), US Department of Energy (DOE) (award no. DE-SC0019375). C.P.N.T. and V.R.K.W. were supported by the NSF (Graduate Research Fellowship no. DGE1106400). A.J. is partially supported by Kwanjeong Educational Foundation. N.S.G. was supported by a David and Lucile Packard Foundation Fellowship for Science and Engineering and Camille and a Henry Dreyfus Teacher-Scholar Award. This work made use of the shared facilities at the University of Chicago Materials Research Science and Engineering Center, supported by National Science Foundation under award number DMR-2011854. Use of the Stanford Synchrotron Radiation Light Source, SLAC National Accelerator Laboratory, is supported by the DOE, Office of Science, Office of Basic Energy Sciences (contract no. DE-AC02-76SF00515). Work performed at the Center for Nanoscale Materials and Advanced Photon Source, U.S. Department of Energy Office of Science User Facilities, was supported by the U.S. DOE, Office of Basic Energy Sciences, under Contract No. DE-AC02-06CH11357.

## ASSOCIATED CONTENT

**Supporting Information**

Additional figures and tables are given in the Supporting Information. This material is available free of charge *via* the Internet at http://pubs.acs.org.

## AUTHOR INFORMATION

**Corresponding Author**

dvtalapin@uchicago.edu

**Notes**

The authors declare no competing financial interests.

Supplementary Information for

# Colloidal dispersions of sterically and electrostatically stabilized PbS quantum dots: the effect of stabilization mechanism on structure factors, second virial coefficients, and film-forming properties

Ahhyun Jeong,[§] Josh Portner,[§] Christian P. N. Tanner, Justin C. Ondry, Chenkun Zhou, Zehan Mi, Youssef A. Tazoui, Vivian R. K. Wall, Naomi S. Ginsberg, and Dmitri V. Talapin*

Corresponding author: Dmitri V. Talapin, dvtalapin@uchicago.edu

[§]These authors contributed equally.

## 1. Materials and Methods

### 1.1 Materials

List of precursors:

Arsenic (V) sulfide (Santa Cruz Biotechnology, 99.99%), cadmium (II) nitrate tetrahydrate (Sigma Aldrich, >99%), hexamethyldisilazane (HMDS, Supelco, 33350-U), hexyl isothiocyanate (Sigma Aldrich, HPLC grade >98%), lead (II) oxide (Sigma Aldrich, 99.999% trace metal basis), myristic acid (Sigma Aldrich, Sigma Grade >99%), *N,N'*-diphenylthiourea (Sigma Aldrich, 98%), *N*-dodecylamine (Sigma Aldrich, 98%), oleic acid (Fisher Scientific, 99%), phenyl isothiocyanate (Sigma Aldrich, 98%), potassium sulfide (Sterm, anhydrous >95%), selenium dioxide (Sigma Aldrich, >99.9% trace metal basis), sodium hydroxide pellets (Sigma Aldrich, 99.99% trace metal basis), sodium stannate trihydrate (Sigma Aldrich, 95%), sodium sulfide nonahydrate (Sigma Aldrich, >98%), trifluoroacetic acid (Fisher Scientific, >99% for spectroscopy), trifluoroacetic anhydride (Sigma Aldrich, 99%), triethylamine (Sigma Aldrich, >99.5%)

List of solvents:

Acetone (Fisher Scientific, >99.5%), acetonitrile (Sigma Aldrich, anhydrous 99.8%), diethylene glycol dimethyl ether (Sigma Aldrich, anhydrous 99.5%), formamide (Sigma Aldrich, >99.5%), hexanes (Fisher Scientific, >98.5%), 1-octadecene (Sigma Aldrich, technical grade 90%), 1-octene (Sigma Aldrich, for synthesis >97%), 2-isopropanol (Fisher Scientific, >99.5%), methanol (Fisher Scientific, >99.9%), methyl acetate (Sigma Aldrich, anhydrous 99.5%), methylcyclohexane (Sigma Aldrich, anhydrous >99%), *N,N*-dimethylformamide (Sigma Aldrich, anhydrous 99.8%), *N*-methylformamide (Sigma Aldrich, 99%), *N*-methylpropionamide (Sigma Aldrich, 98%), *n*-hexane (Sigma Aldrich, anhydrous 95%), toluene (Fisher Scientific, >99.8%)

### 1.2 Synthesis of quantum dots (QDs)

*PbS QDs:* The synthesis of PbS QDs follows the previously reported method.[1] After washing, the samples were stored in $N_2$ atmosphere in anhydrous methylcyclohexane or anhydrous *n*-hexane.

*CdSe QDs:* The synthesis of CdSe QD follows the previously reported method.[2] After washing, the samples were stored in methylcyclohexane or hexanes.

### 1.3 Preparation of inorganic ligands

The preparation of $K_4Sn_2S_6$ and $K_3AsS_4$ ligands follows previously reported procedure.[3]

### 1.4 Ligand exchange of organic to MCC ligands

The bi-phasic ligand exchange of organically capped QDs is adapted from a previously reported procedure.[3] In a typical ligand exchange, 4 ml of 25 mM $K_4Sn_2S_6$ in distilled NMF, 8-10 ml of anhydrous hexane and approximately 100 mg of PbS-OA QDs were stirred in a 20 ml vial for ~6 hours using a PTFE-coated stir bar. After stirring, the PbS QDs completely transferred from the top hexane layer to the bottom NMF layer. The hexane layer was subsequently replaced with fresh hexane three times, with 30 minutes stirring between each replacement. Then, PbS-$Sn_2S_6$ QDs in distilled NMF layer were precipitated by adding acetonitrile as a non-solvent. The resulting PbS-$Sn_2S_6$ QDs were redispersed in 4 ml of 25 mM $K_4Sn_2S_6$ in distilled NMF and stirred with 8-10 ml of fresh hexane. The hexane layer was replaced with fresh hexane three times. The PbS-$Sn_2S_6$ QDs were then precipitated using acetonitrile and subsequently redispersed in distilled NMF. Finally, the solution was filtered using a 0.22 μm PVDF membrane syringe filter and stored under an $N_2$ atmosphere. All the procedures were carried out in a dry $N_2$ glovebox.

### 1.5 Film preparation

For SEM measurements, the 525 μm thick nitrogen doped Si wafer with 300 nm layer of dry-grown thermal oxide was cleaned by sonicating successively in hexane, acetone, DI water and isopropanol for 5 minutes each. To prepare the films for PbS-OA QDs, the surface of the wafer was treated by spin-coating HMDS at 3000 rpm for 45 seconds, followed by annealing at 150°C for 30 minutes. A solution of PbS-OA in MCH was drop-casted onto the wafer and the solvent was allowed to evaporate under ambient pressure and temperature. During the evaporation, the wafer was tilted by ~10° by placing one edge on a glass microscope slide. To prepare the films for PbS-$Sn_2S_6^{4-}$ QDs, the same cleaning method was used, and the surface of the wafer was treated with oxygen plasma (Harrick Plasma, Plasma Cleaner PDC-001) for 15 minutes. A solution of PbS-$Sn_2S_6^{4-}$ in NMF was drop-cast onto the wafer and the solvent was allowed to evaporate under vacuum. The wafer was similarly tilted by ~10° during the evaporation process.

For SAXS measurements, the solution of QDs were drop-cast onto a boron-doped Si wafer with a thickness of 50-75 μm. MCH was allowed to evaporate under ambient temperature and pressure, and NMF was allowed to evaporate under vacuum.

### 1.6 Characterization methods

**UV-Vis-NIR Absorption:** UV-Vis-NIR absorption spectra were collected on a Shimadzu UV3600 Plus spectrophotometer.

**TEM:** Transmission electron microscopy (TEM) images were collected using FEI Tecnai G2 F30 X-TWIN at an accelerating voltage of 300 kV. Samples were prepared by drop-casting a dilute solution of QDs on 400 mesh copper grids with amorphous carbon support (Ted Pella).

**SEM:** Scanning electron microscopy (SEM) images were collected using a Zeiss Merlin scanning electron microscope with 10 kV accelerating voltage.

**ICP-OES:** The concentration of QDs were obtained from inductively coupled plasma optical emission spectroscopy (ICP-OES) measurements, which were collected using an Agilent 700 Series instrument or Agilent 5110 Synchronous Vertical Dual View (SVDV). The QD and ligand solutions were digested by previously established protocol.[4]

**Synchrotron SAXS:** Most small-angle X-ray (SAXS) data in figures 1, 2, 3 and 4 are collected at beamline 12-ID-B, Advanced Photon Source (APS) in Argonne National Laboratory. The beam energy was 12.7 keV and the beam size was ~0.5 mm. Colloidal QDs were measured in glass capillaries that were flame-sealed in air. Films of QDs were prepared and measured on a 50-75 μm thick Si wafer.

**Cu K alpha source SAXS:** Few SAXS datasets in figure 1D (SAXS intensity of PbS-OA), figure 2A (structure factor of PbS-OA) and figure 4A (SAXS intensity of PbS-$Sn_2S_6$ with $K_3AsS_4$ film) were collected using a SAXSLAB Ganesha Instrument with a Cu K-alpha source. Colloidal QD solutions were contained in glass capillaries that were flame-sealed in air. Films of QDs were prepared and measured on boron-doped 50-75 μm thick Si wafers.

**XPCS:** X-ray photon correlation spectroscopy (XPCS) data were collected at the Materials Imaging and Dynamics (MID) instrument at the European X-ray Free Electron Laser (XFEL). The X-ray photon energy was 12.3 keV and the sample-to-detector distance was 7 m. Samples were prepared in a 0.9 mm diameter glass capillary. The XFEL produced trains of ultrafast spatially coherent X-ray pulses, where each train consisted of 50 pulses at a rate of 2.2 MHz. The first pulse of each train is delivered 0.1 s after the first pulse of the previous pulse train.

## 2. Characterization of Nanocrystals

### 2.1 UV-Vis-NIR absorption

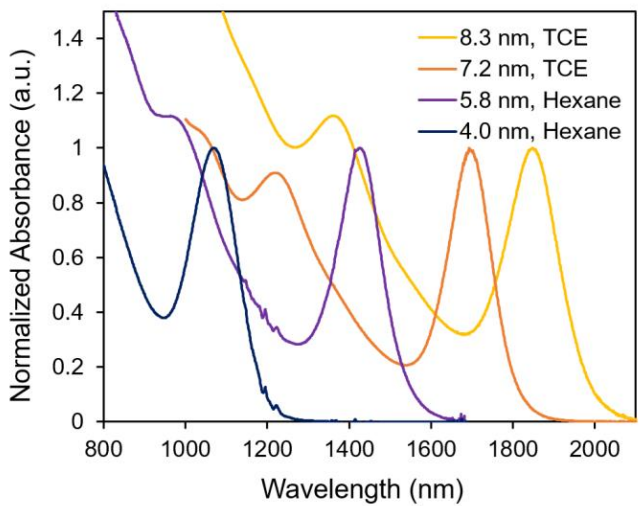

**Figure S1.** The UV-Vis-NIR absorption spectra of PbS-OA QDs in non-polar solvents.

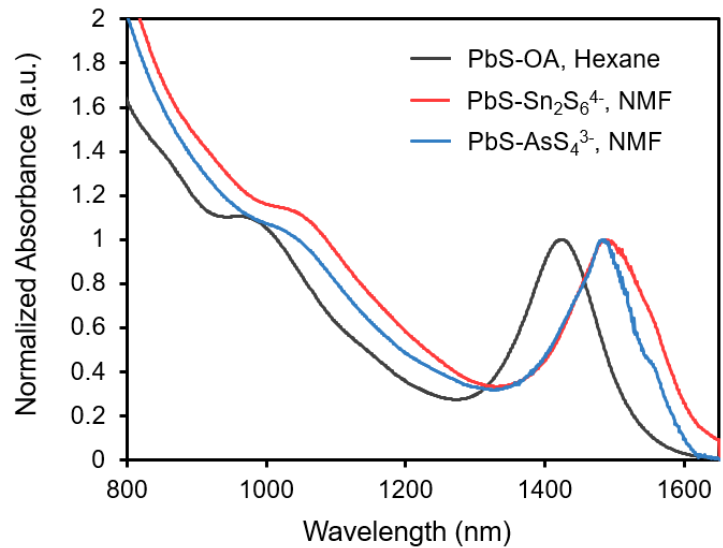

**Figure S2.** The UV-Vis-NIR absorption spectra of 5.8 nm PbS QDs with organic and inorganic ligands. The spectrum red shifts after the ligand exchange, due to the relaxation of the quantum confinement.

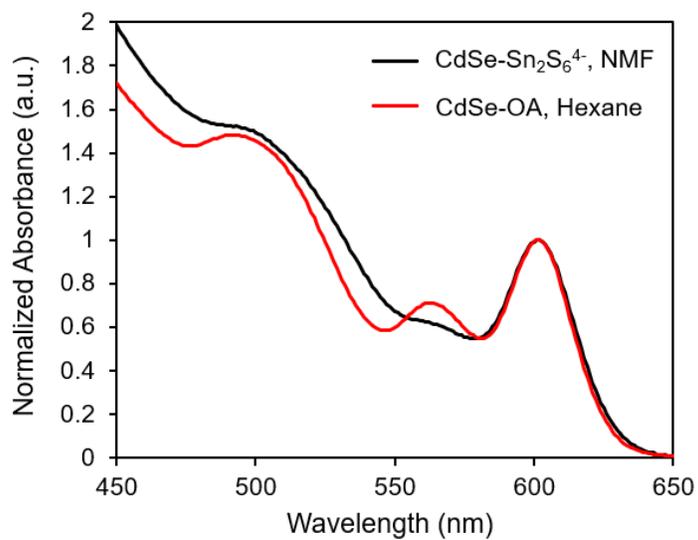

**Figure S3.** UV-Vis spectra of 5.2 nm CdSe-$Sn_2S_6^{4-}$ and CdSe-OA QDs.

S-6

## 2.2 TEM images

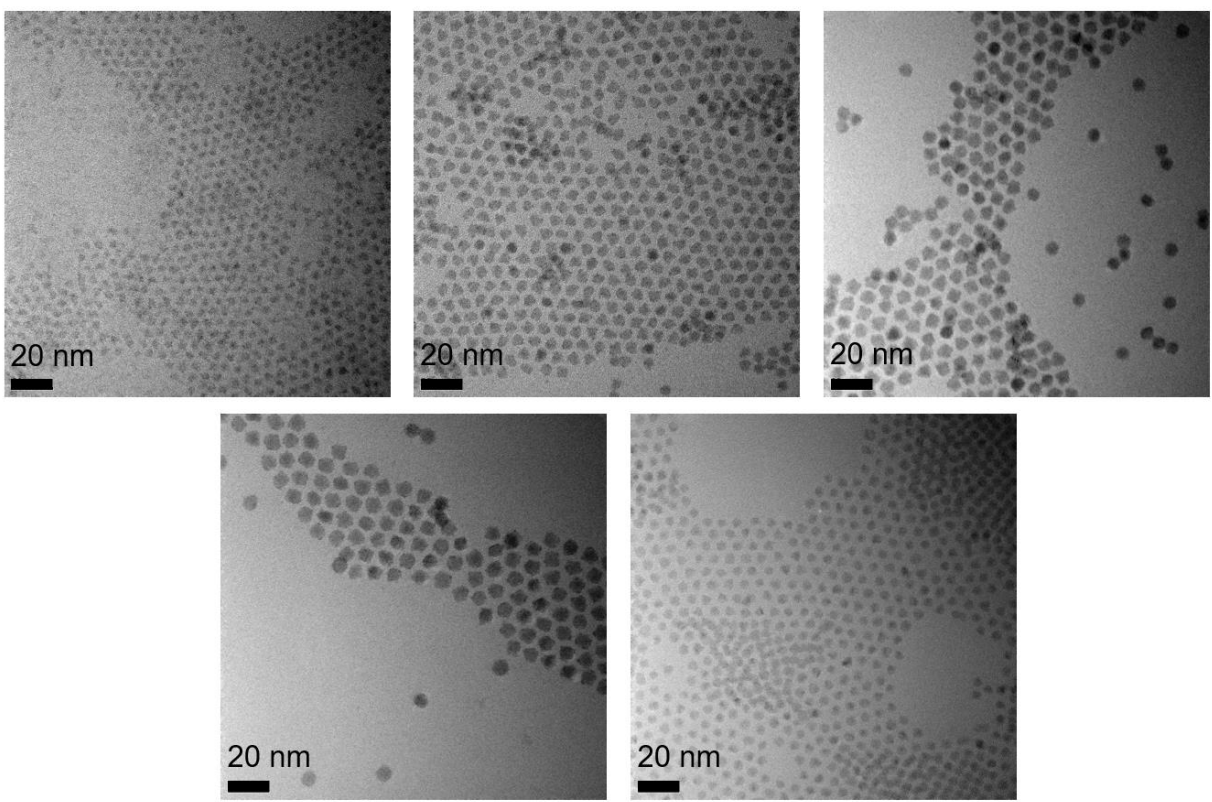

**Figure S4.** TEM images of 4.1 nm PbS-OA (top left), 5.8 nm PbS-OA (top center), 7.2 nm PbS-OA (top right), 8.1 nm PbS-OA (bottom left), and 5.2 nm CdSe-OA (bottom right) QDs.

## 2.3 SAXS patterns and size distribution

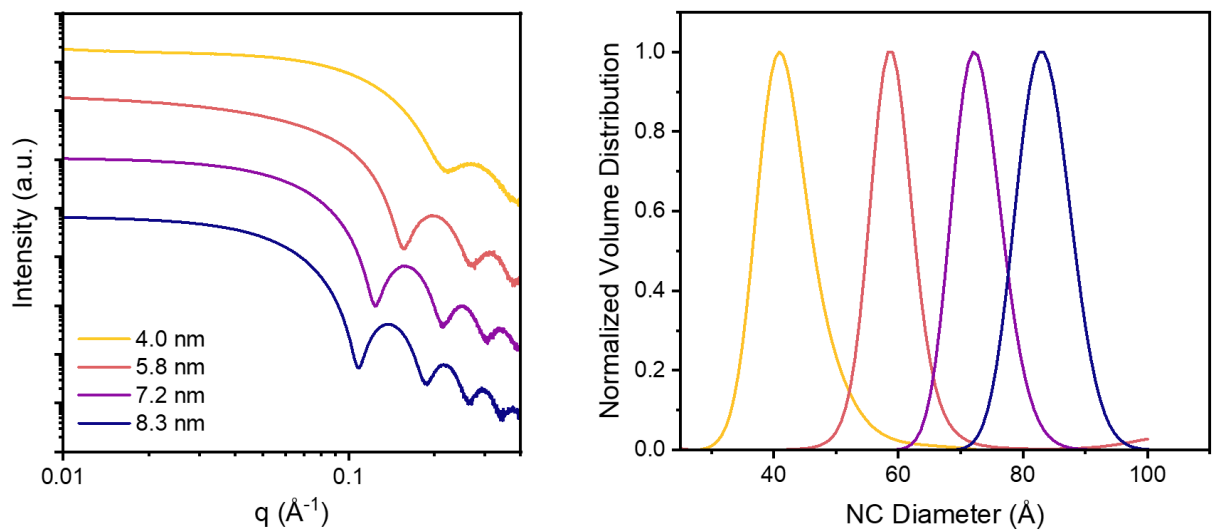

**Figure S5.** SAXS patterns (left) and size distributions (right) of 10 mg/ml PbS-OA QDs in MCH. The size distribution was obtained by analyzing the form factor of SAXS pattern.

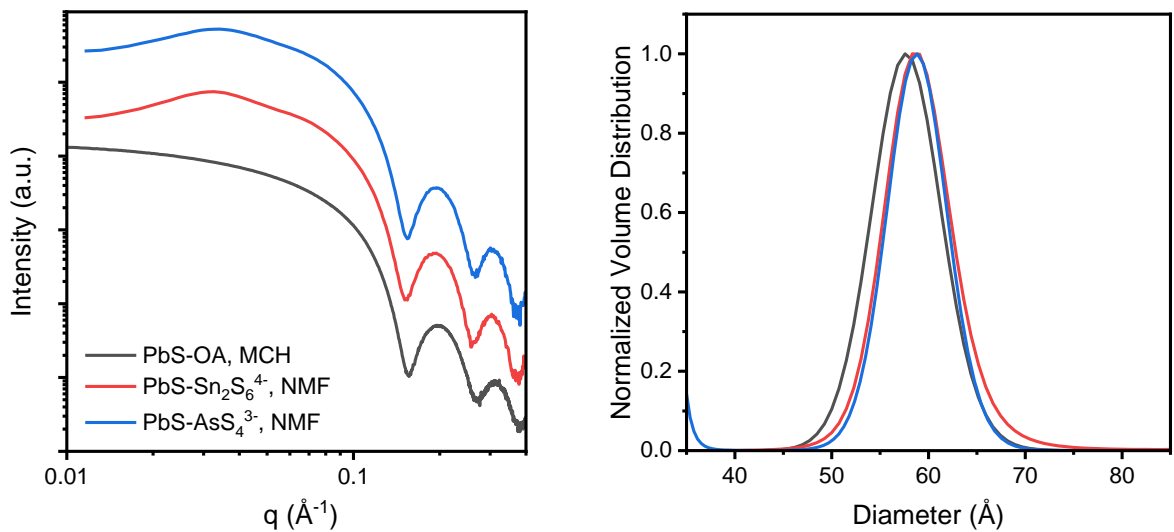

**Figure S6.** SAXS patterns (left) and size distributions (right) of 10 mg/ml 5.8 nm PbS QDs with OA, $Sn_2S_6^{4-}$ and $AsS_4^{3-}$ ligands. The PbS QDs with inorganic ligands appear larger because the heavy elements in the ligands scatter the X-rays.

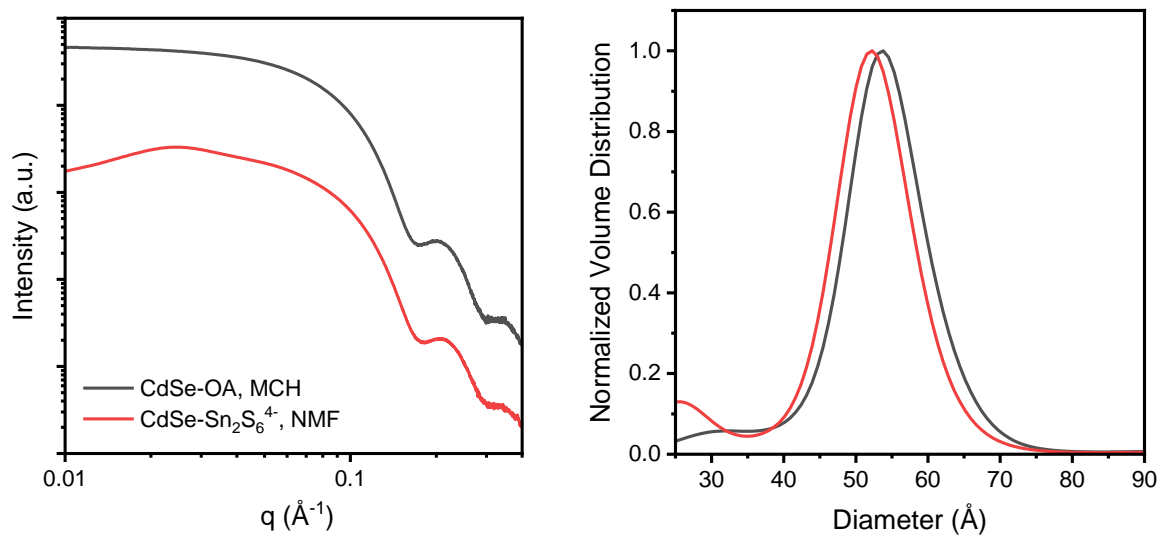

**Figure S7.** SAXS patterns (left) and size distributions (right) of 10 mg/ml 5.2 nm CdSe QDs with OA (black) and $Sn_2S_6^{4-}$ (red) ligands.

## 2.4 Colloidal stability of PbS-$Sn_2S_6^{4-}$ QDs in different polar solvents

Although QDs may appear well-dispersed to the naked eye, their colloidal stability can be lacking. SAXS measurement is a reliable method to evaluate the colloidal stability of QDs. When QDs are colloidally unstable, the QDs tend to undergo micro-aggregation within the solution, resulting in an up-turn in the low-q region (Figure S8). This scattering behavior follows the following equation, where $I(q)$ denotes the SAXS intensity, $q$ represents the scattering momentum and $n$ represents the dimensionality of the aggregates.[5] For example, an up-turn with $n \approx 4$ indicates the presence of three-dimensional aggregates in an QD solution.

$$I(q) \propto q^{-n}$$

Examples of such behavior can be observed in Figure S8, illustrating PbS-$Sn_2S_6^{4-}$ QDs with poor colloidal stability. Despite initially appearing well-dispersed in solution, these colloids may gradually aggregate and eventually precipitate into solids over time. Conversely, a colloidally stable solution of QDs shows no rise of intensity in the low-q region of SAXS, as depicted in Figure 1F in the main text.

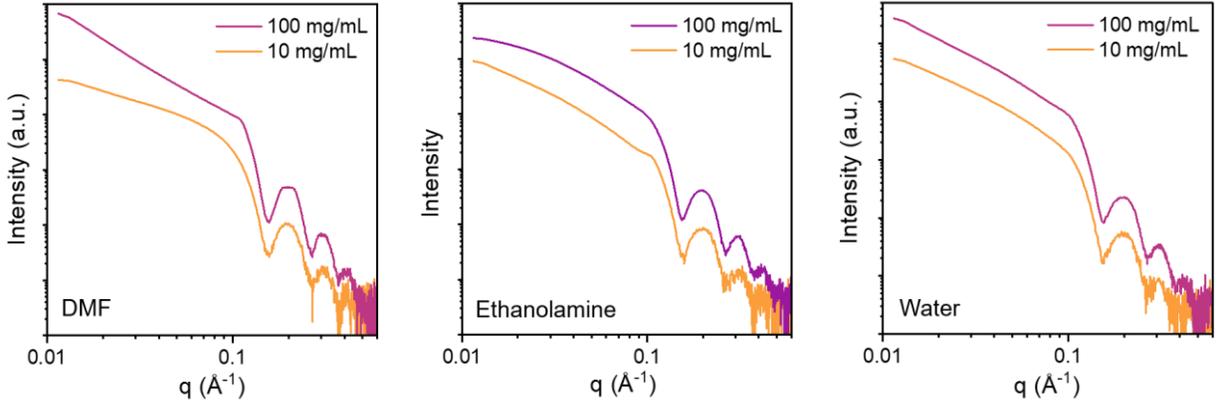

**Figure S8.** The SAXS patterns of 5.7 nm PbS-Sn$_2$S$_6^{4-}$ QDs in DMF (left), ethanolamine (center) and water (right).

### 3. SAXS analysis of colloidal QDs

#### 3.1 Separation of SAXS to form factor and structure factor

The SAXS intensity ($I(q)$) is a product of a scaling factor ($A$), form factor intensity ($P(q)$) and structure factor ($S(q)$).

$$I(q) = A\, P(q)\, S(q)$$

The form factor intensity can be modelled as the following, where $F(q, R)$ is the form factor amplitude, $n(R)$ is the size distribution of QDs, $q$ is scattering vector and $R$ is the radius of the QDs.

$$P(q) = \int n(R)\, \langle |F(q, R)|^2 \rangle\, dR$$

$$F(q, R) = \frac{3\big(\sin(qR) - qR\cos(qR)\big)}{(qR)^3}$$

The size distributions of the QDs were determined by fitting the Porod region of the background-subtracted SAXS intensity with Irena tool suite in Igor Pro. The fitting was performed using the maximum entropy method. The form factor intensity was then fitted with a Gaussian function to obtain the size distribution by number fraction. This Gaussian function was substituted into $n(R)$ to obtain the form factor. The scaling factor ($A$) was determined by fitting the form factor intensity with the SAXS intensity at the first Bessel peak ($4.493 < qR < 7.725$) using the least squares method. The resulting fits of form factor and SAXS intensity are shown in Figure S9. The structure factor was calculated by dividing the scaling factor and form factor intensity from the SAXS intensity.

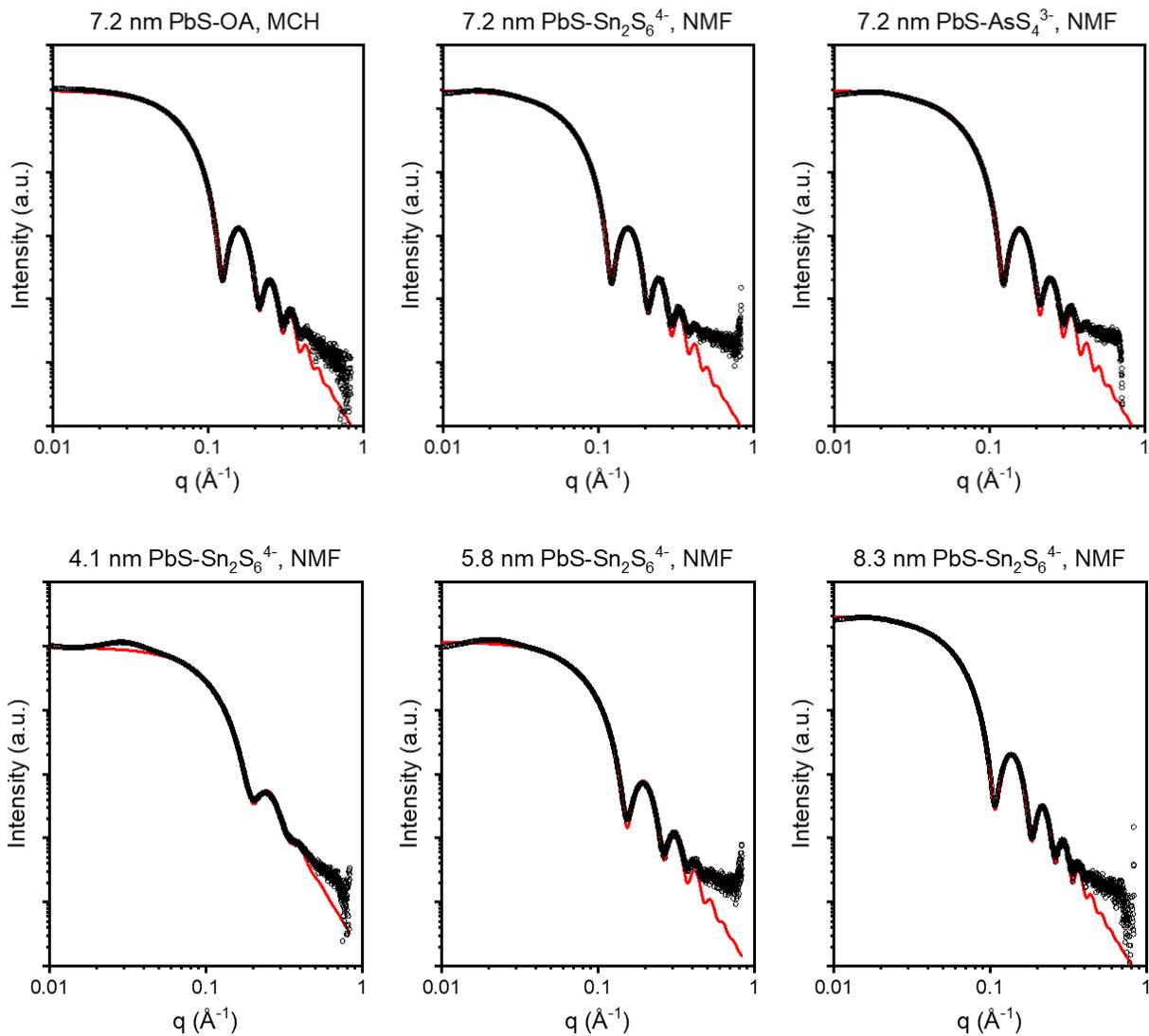

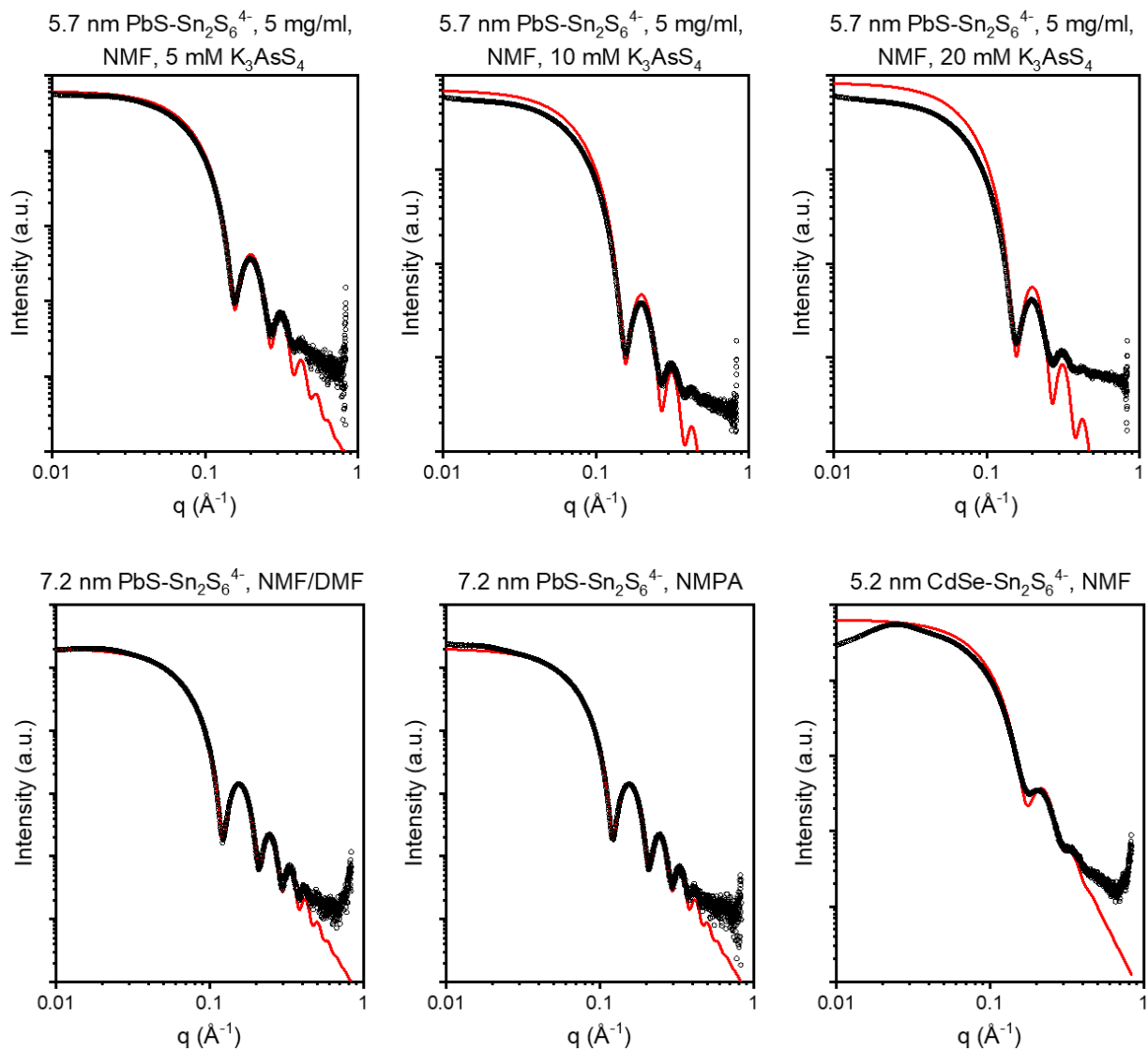

**Figure S9.** Background subtracted SAXS patterns (black circles) and fitted form factors (red lines). All QDs are in concentrations of 10 mg/ml, unless otherwise specified.

### 3.2 Size of QDs

All the reported values of the sizes of the QDs are calculated using the following procedure. Initially, the size distribution of oleate-capped QDs was obtained by fitting the Porod region of SAXS using IgorPro9. Subsequently, the size distribution obtained was fitted with a Gaussian function, and the diameter at the peak of this Gaussian distribution was reported as the size of QDs. In this article, the reported sizes of all QDs, regardless of the capping ligand, were obtained from the oleate-capped QDs. This is because organic molecules such as oleate are mostly X-ray transparent, enabling a more accurate determination of the size of the core-only QD. On the other hand, inorganic ligands such as $Sn_2S_6^{4-}$ or $AsS_4^{3-}$ ligands strongly scatter X-ray, making the QDs appear larger in the SAXS data.

## 3.3 Percus-Yevick approximation

The structure factor of hard spheres can be derived by solving the Ornstein-Zernike equation using the Percus-Yevick approximation.[6] By fitting the experimental structure factor with the analytical solution, we can extract information regarding the radius and volume fraction of colloidal hard spheres with a similar structure factor. By comparing the fitted radius and volume fraction, we can determine how closely our system resembles the hard spheres. The analytical solution for the structure factor of hard spheres ($S_{PY}(q)$) is given as follows, where $A = 2qR$, $\eta$ represents the volume fraction and $R$ is the radius of the hard sphere. The characteristic center-to-center distance between neighboring hard spheres ($D_{NC-NC}$) is equal to $2R$.

$$S_{PY}(q) = \frac{(1+2\eta)^2}{A^2(1-\eta)^4}(\sin A - A\cos A) - \frac{6\eta\left(1+\frac{\eta}{2}\right)}{A^3(1-\eta)^4}(2A\sin A + (2-A^2)\cos A - 2)$$
$$+ \frac{\eta(1+2\eta)^2}{2A^5(1-\eta)^4}\left((-A^4 + 12A^2 - 24)\cos A + 4A(A^2-6)\sin A + 24\right)$$

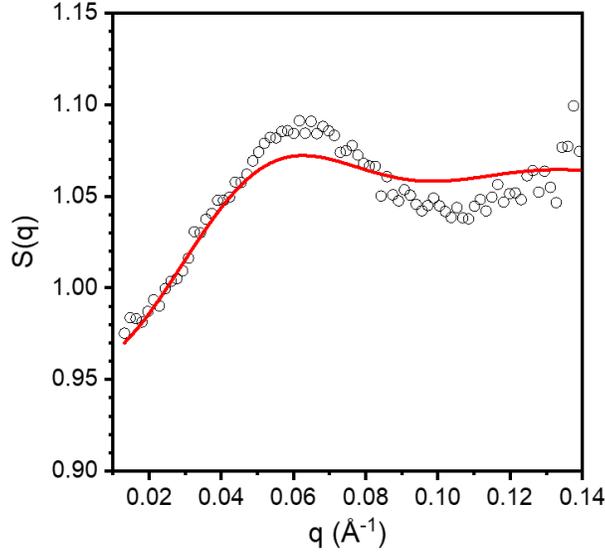

**Figure S10.** Experimental structure factor (black circle) of 5.7 nm PbS-OA and the corresponding fits with the analytical structure factor of the hard sphere model (red line). The concentration of the QDs is 50 mg/ml.

Figure S10 presents the experimental structure factors of PbS-OA QDs, along with the corresponding fits using the analytical structure factors of the hard sphere model. By fitting the structure factor, we find that the PbS-OA QDs can be modelled as hard spheres with the diameter ($2R$) and volume fraction ($\eta$) of 92 Å and 1.3%, respectively. This is in good agreement with the actual diameter and volume fraction of the PbS-OA QDs, which are approximately 87 Å and 2.3%, respectively, including the ligands.

### 3.4 Characteristic distance between QDs from $q_{max}$

The characteristic center-to-center distance ($D_{NC-NC}$) between two neighboring QDs was calculated using the following equation, where $q_{max}$ is value of scattering vector ($q$) at the first peak of the structure factor.

$$D_{NC-NC} = \frac{2\pi}{q_{max}}$$

The log-log plot of $q_{max}$ and $D_{NC-NC}$ against the concentration of 5.7 nm PbS-Sn$_2$S$_6^{4-}$ QDs is shown in Figure S11. A linear fit of the log-log plot of $D_{NC-NC}$ against QD concentration yields the following empirical relationship, where $c$ is the concentration of QD in mg/ml.

$$D_{NC-NC} = 488\, c^{-0.267}$$

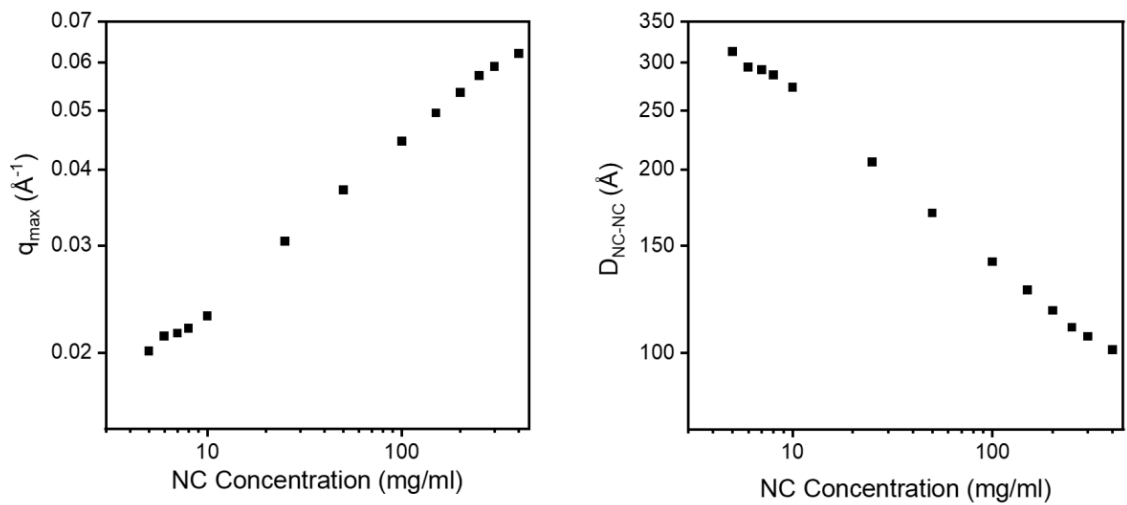

**Figure S11.** Log-log plot of q$_{max}$ against QD concentration (left) and $D_{NC-NC}$ against QD concentration (right). The QDs are 5.7 nm PbS-Sn$_2$S$_6^{4-}$.

### 3.5 Structure factor of PbS-Sn$_2$S$_6^{4-}$ QDs in NMPA

The structure factor of PbS-Sn$_2$S$_6^{4-}$ QDs in NMPA were also obtained using the method described in Section 3.1. The shift of $q_{max}$ in respect to the number density of the QDs is shown and analyzed in the Figure 2C of the main text.

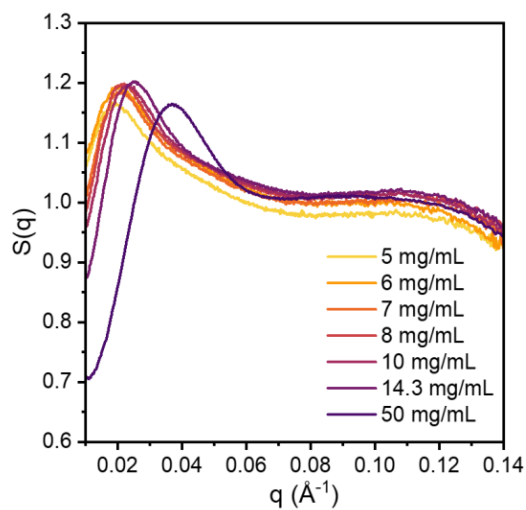

**Figure S12.** Concentration-dependent structure factors of 5.7 nm PbS-$Sn_2S_6^{4-}$ QDs in NMPA.

### 3.5 Hyperuniformity index

Hyperuniformity is a property found in certain materials where density fluctuations are suppressed over large length scales. Examples of hyperuniform material include perfect crystals, jammed hard spheres and particles separated by long-range soft potential. Disordered hyperuniform materials have been of particular interest in scientific community due to their resistance to defects and thermal fluctuations, as well as their isotropic optical properties. These materials hold potential applications in fields like waveguides and as sources of isotropic thermal radiation.[7]

The hyperuniformity of a material can be evaluated through the analysis of its structure factor. Key characteristics of disordered hyperuniform material are identified as follows: (i) a prominent primary peak in the structure factor indicates a uniform separation between particles, and (ii) the structure factor approaches 0 at $q = 0$, indicating minimal density fluctuations at long length scales. The hyperuniformity index ($h$) serves as a measure of uniformity within a material. It is computed by dividing the value of structure factor at the primary peak ($S_{max}$) by the structure factor at $q = 0$, where $q$ represents the scattering vector.

$$h = \frac{S(0)}{S_{max}}$$

A perfectly hyperuniform material has a hyperuniformity index of 0, while a fully randomly arranged material, like an ideal gas, has a hyperuniformity index of 1. Materials with $h < 10^{-3}$ are typically classified as "effectively hyperuniform", indicating significant suppression of density fluctuations over large scales.[8]

The hyperuniformity index of 50 mg/ml PbS-OA QDs in hexane and PbS-$Sn_2S_6^{4-}$ QDs in NMF is 0.89 and 0.39 respectively, suggesting that the electrostatically stabilized QDs in a polar solvent exhibit a significantly more uniform arrangement compared to that of sterically stabilized QDs in a non-polar solvent. In Figure S13, the hyperuniformity index of 5.7 nm PbS-$Sn_2S_6^{4-}$ QDs in NMF is shown at a series of concentrations. It is observed that the hyperuniformity index decreases monotonically with increasing concentration, indicating at the arrangement of QDs in a concentrated solution is much more uniform compared to that in a dilute solution.

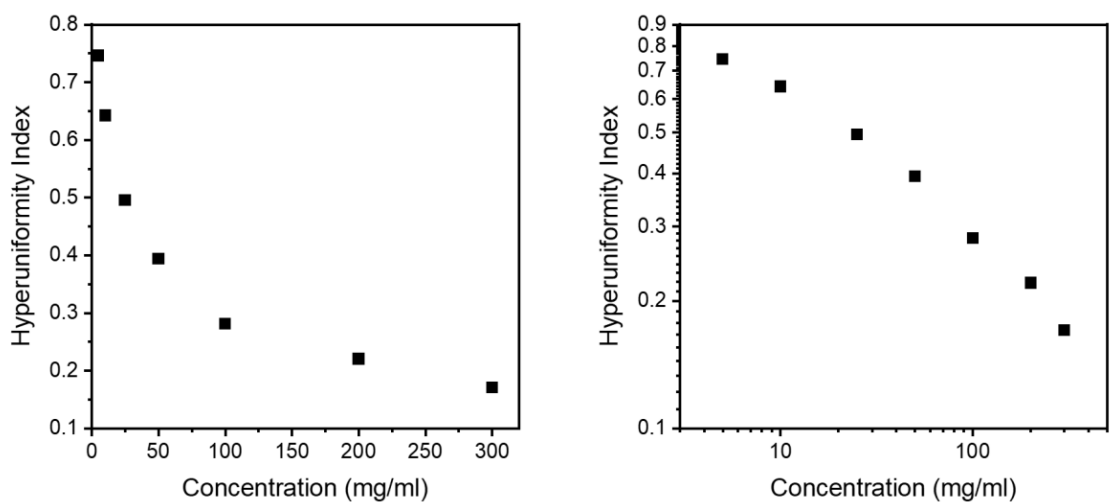

**Figure S13.** Lin-lin plot (left) and log-log plot of the hyperuniformity index of 5.7 nm PbS-$Sn_2S_6^{4-}$ QDs in NMF against concentrations.

### 3.6 Diffusion coefficients of PbS-$Sn_2S_6^{4-}$ QDs

The diffusion coefficient of PbS-$Sn_2S_6^{4-}$ QDs at concentrations of 21.6 mg/mL and 400 mg/mL dissolved in 60:40 and 100:0 ratios of N-methyl formamide to *N,N*-dimethylformamide, respectively, were prepared and measured with XPCS. Two-dimensional X-ray scattering data were collected. The analysis of the XPCS data follows previously reported literature.[9]

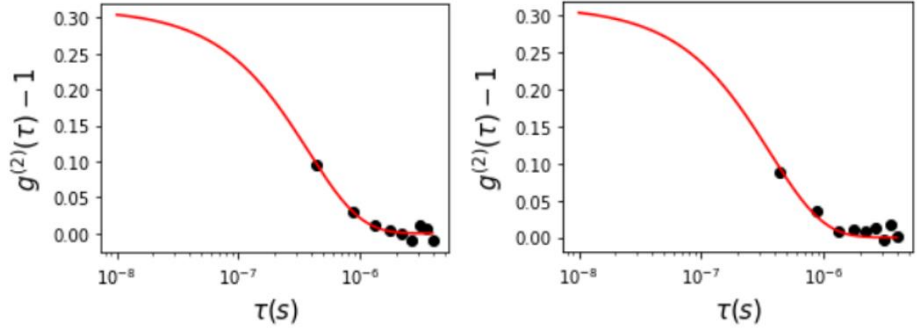

**Figure S14.** Experimentally obtained autocorrelation (black circles) and fit (red line) of 21.6 mg/ml (bottom, left) and 400 mg/ml (bottom, right) 5.8 nm PbS-$Sn_2S_6^{4-}$ QDs at $q = 0.016 Å^{-1}$. Two-time correlation functions are defined as the following:

$$C(t, t+\tau) = \frac{\langle \delta I(t) \, \delta I(t+\tau) \rangle_{sp}}{\langle I(t) \rangle_{sp} \langle I(t+\tau) \rangle_{sp}}$$

where $\delta I(t) = I(t) - \langle I(t) \rangle_{sp}$, and $\langle I(t) \rangle_{sp}$ denotes an average over speckles in a region of interest on the detector centered at $q = 0.016 Å^{-1}$.[10] The two-time correlation functions were calculated over the 10 pulses within each pulse train and then averaged across all pulse trains for each sample. The intensity-intensity autocorrelation ($g_2(\tau)$) is calculated from $C(t, t+\tau)$ by averaging over times ($t$):

$$g_2(\tau) - 1 = \overline{C(t, t+\tau)}$$

The autocorrelation functions are fit to the following equation:

$$g_2(\tau) - 1 = \beta \exp(-2\gamma\tau)$$

where $\beta$ is the speckle contrast and $\gamma$ is the rate at which the autocorrelation function decays. $\beta$ in the fits in Figure S14 (left, right) is fixed at 0.312 based on the average of the direct calculation of the speckle contrast, i.e., $C(t, t)$, of hundreds of individual images. The decorrelation rate ($\gamma$) is related to the effective diffusion coefficient ($D_{eff}$) of the QDs via $\gamma = D_{eff} q^2$, where $q$ is the scattering vector that is dependent on the wavelength of the X-ray and the scattering angle. For our measurements, the value of $q$ was fixed at 0.016 Å$^{-1}$. From the exponential fits, we extract $D_{eff} = 47 \pm 5 \, \mu m^2/s$ for the 21.6 mg/mL sample and $D_{eff} = 51 \pm 5 \, \mu m^2/s$ for the 400 mg/mL sample.

Using the Stokes-Einstein relation:

$$D_{\text{eff}} = \frac{k_B T}{3\pi\eta d_H}$$

Where $k_B$ is Boltzmann's constant, T is the temperature, $\eta$ is the solvent viscosity, and $d_H$ is the hydrodynamic diameter of the QDs, we extract effective diffusion coefficients corresponding to hydrodynamic diameters ~ 6-7 nm.

## 4. B₂ coefficients

### 4.1 Experimental determination of B₂ coefficients

The second virial ($B_2$) coefficients were determined experimentally from concentration-dependent SAXS measurements. Initially, solutions of QDs with various concentrations were prepared by measuring the concentration of the stock QD solution using ICP-OES, and then diluting it to the desired concentration. The QD solutions were transferred to a glass capillary for SAXS measurements. Next, the structure factor was extracted from the SAXS patterns of the QDs using the procedure outlined in Section 3.1. For the electrostatically stabilized QDs, solutions with concentrations ranging from 3 to 14 mg/ml were measured for the analysis. For the sterically stabilized PbS-OA QDs, SAXS measurements were taken at higher concentrations (25-300 mg/ml) because the structure factor was too small to be analyzed at lower concentrations. Figure 3a in the main text presents a representative structure factor from 7.2 nm PbS-Sn$_2$S$_6^{4-}$ QDs in NMF, and the structure factors of all other analyzed QDs are shown in Section 4.2.

The $B_2$ coefficient and the value of structure factor at $q = 0$ ($S(0)$) are related by the following equation, where $n$ represents the number density of the QDs in the solution. By rearranging the equation, we can estimate the $B_2$ coefficient by halving the slope of the plot of $1/S(0) - 1$ against the number density ($n$).

$$\frac{1}{S(0)} = 1 + 2B_2 n + \mathcal{O}(n^2)$$

The number density ($n$) can be calculated from the following equation, where $c$ is the concentration of QDs in mg/ml, $\rho$ is the density of QDs and $V_{NC}$ is the volume of the QDs.

$$n = \frac{c}{\rho V_{NC}}$$

The values of $S(0)$ at each concentration of QD solution were determined by extrapolating the experimentally determined structure factor. To achieve this, the structure factor was plotted as $\ln[S(q)]$ against $q^2$. In the low-q region, the plot of $\ln[S(q)]$ against $q^2$ shows linear relationship (see Figure S15). This region of the plot was fitted linearly to extrapolate the structure factor at $q = 0$. Figure 3b in the main text is a representative plot of $1/S(0) - 1$ against the number density from 7.2 nm PbS-$Sn_2S_6^{4-}$ QDs in NMF, and the plot of all other QDs are shown in figure S16.

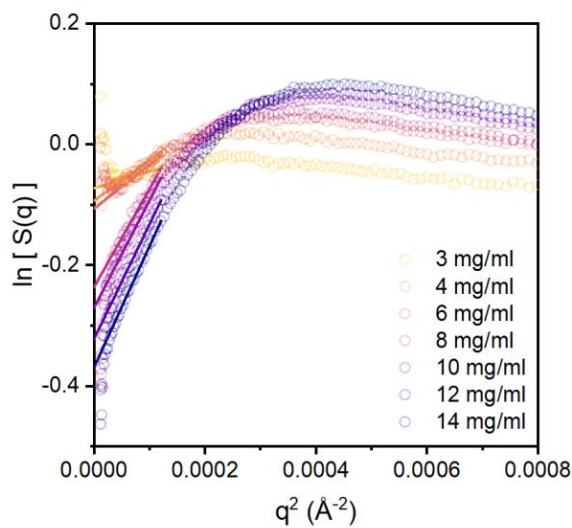

**Figure S15.** A representative plot of $\ln[S(q)]$ against $q^2$. This plot was obtained from the experimentally determined structure factor of 7.2 nm PbS-$Sn_2S_6^{4-}$ QDs in NMF.

## 4.2 Structure factor and the analysis of B$_2$ coefficient for different PbS QD samples

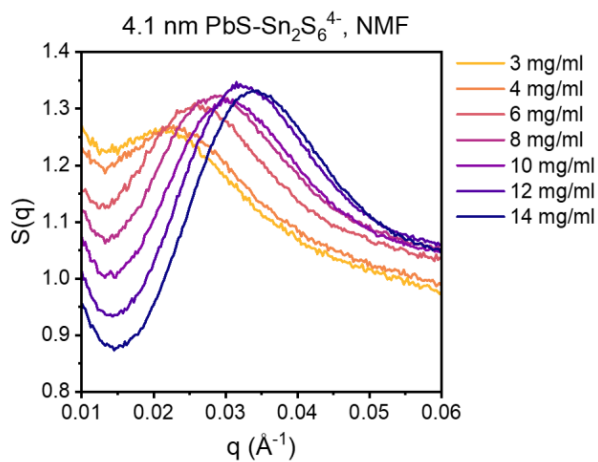
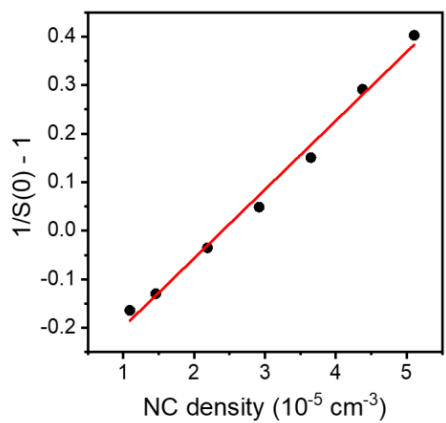

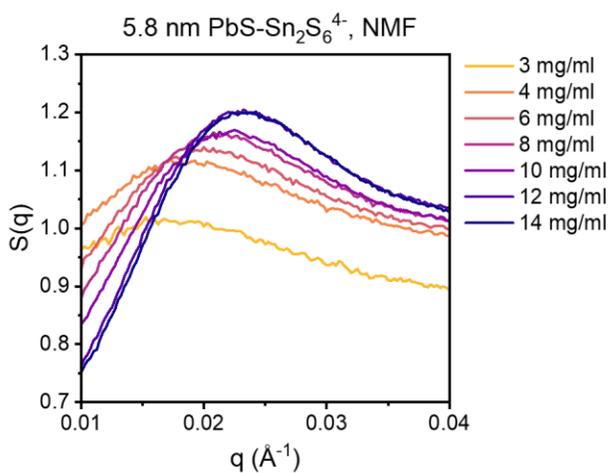
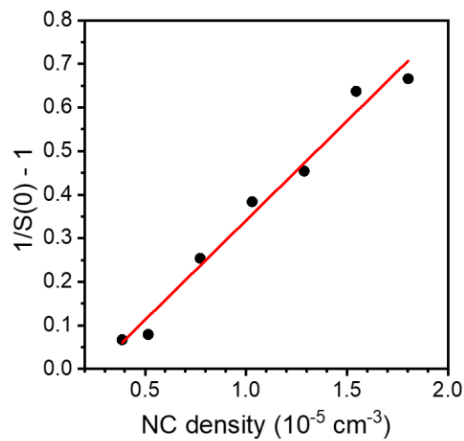

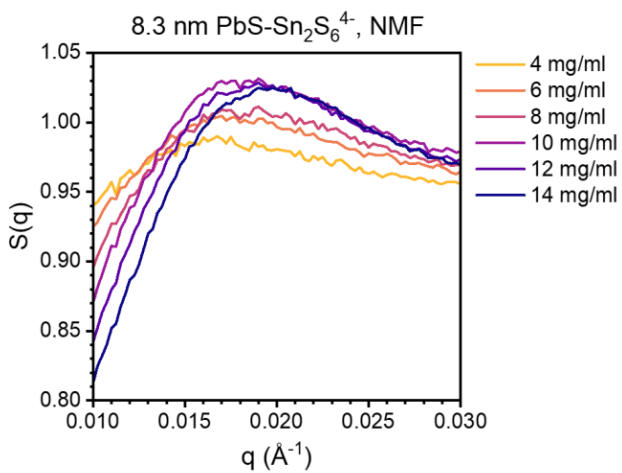
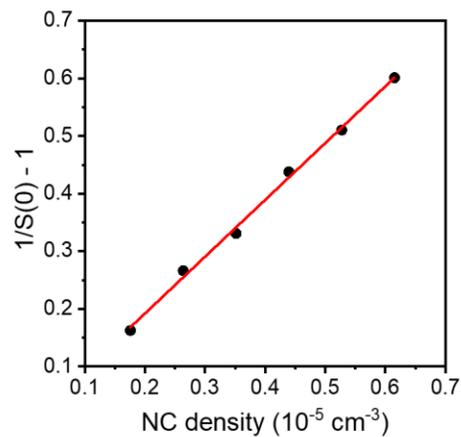

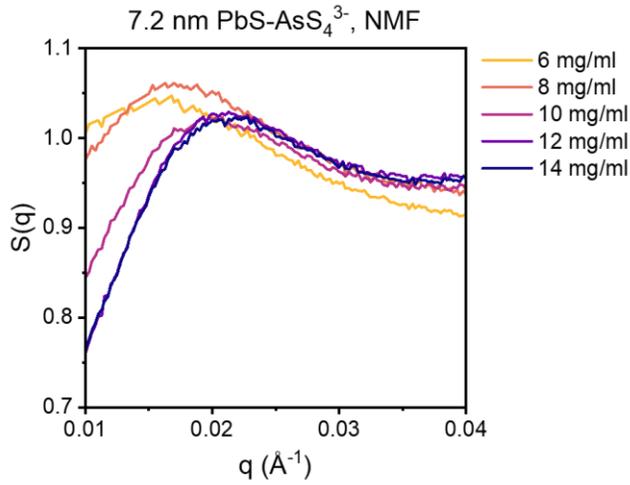
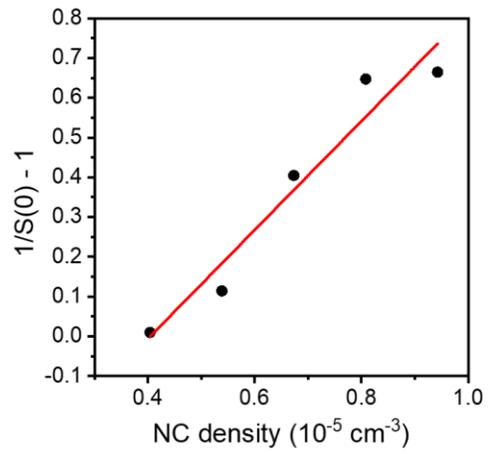
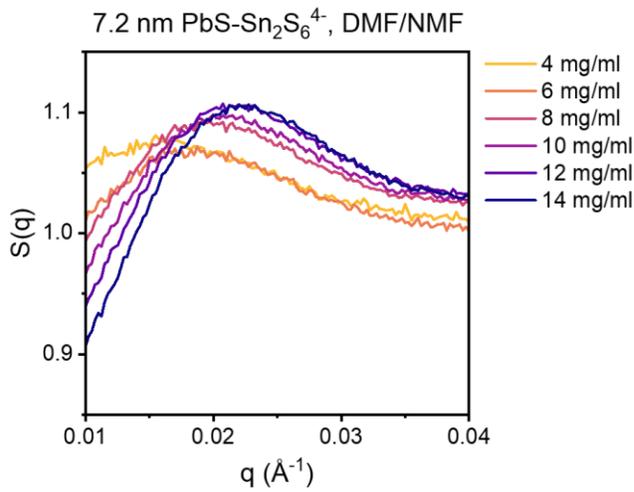
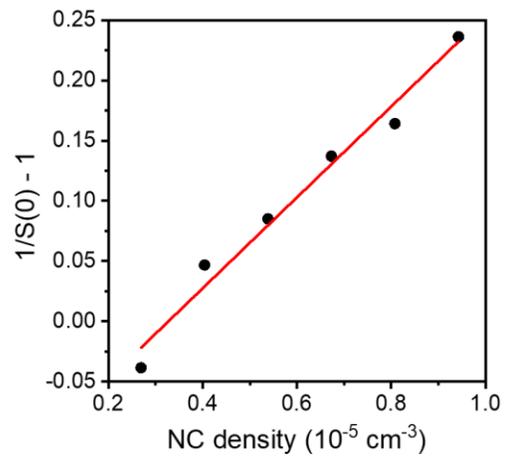
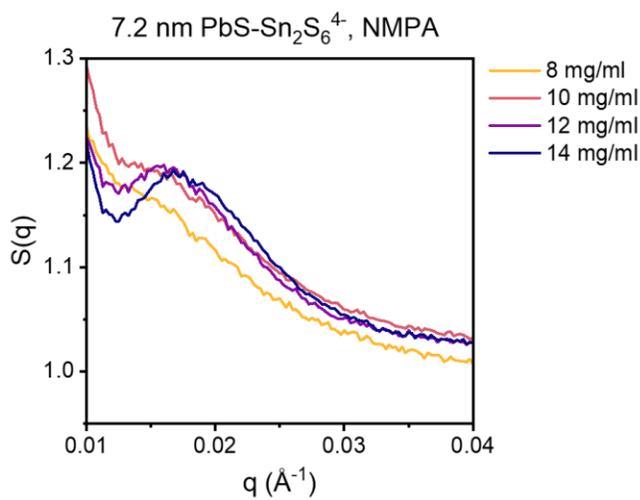
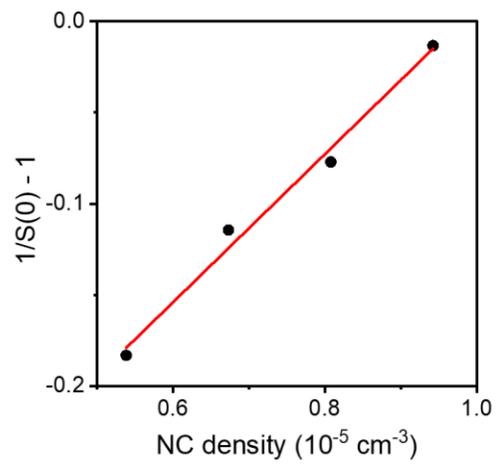

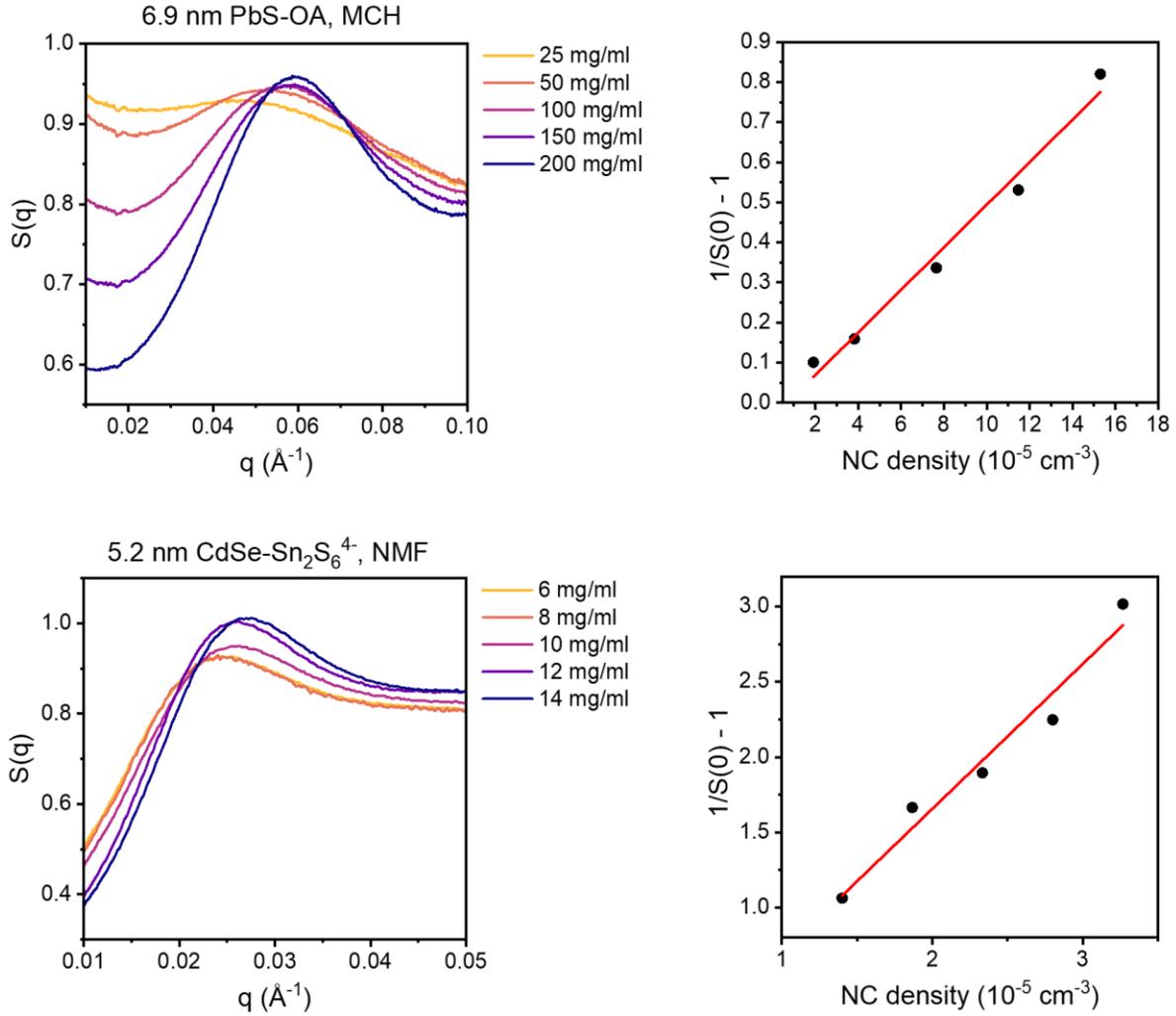

**Figure S16.** Plots of the structure factors (left) and the values of 1/S(0)-1 against the number density of QDs (right).

### 4.3 Determination of the QD size for calculating $b_2$ coefficient

$b_2$ coefficients were calculated by $b_2 = B_2/B_{2,HS}$, where $B_{2,HS} = 2\pi(d+2\delta)^3/3$. In this equation, $d$ is the diameter of the core-only QD and $\delta$ is the length of the ligand. The diameter of the core-only QD was experimentally obtained through the process described in Section 3.2. The length of the ligands were estimated by subtracting the diameter of the core-only QDs from the nearest-neighboring distance of superlattice of the QDs (see Section 7.1). The length of the oleate (OA) ligand is estimated to be 1.1 nm. The length of the $Sn_2S_6^{4-}$ and $AsS_4^{3-}$ ligands are estimated to be 0.29 nm and 0.13 nm, respectively. These values are in line with values from the previous literature.[11-13]

### 4.4. B₂ coefficient of CdSe QDs

Further analysis of 5.2 nm CdSe-Sn$_2$S$_6^{4-}$ QDs revealed that the $B_2$ and $b_2$ coefficient of these QDs are $B_2 = 6.28 \times 10^7$ Å$^3$ and $b_2 = 117$, which indicates that the CdSe-Sn$_2$S$_6^{4-}$ QDs exhibit significantly stronger interparticle repulsion than PbS-Sn$_2$S$_6^{4-}$ QDs. Several factors potentially contribute to this discrepancy. First, the PbS-Sn$_2$S$_6^{4-}$ (7.2 nm) and CdSe-Sn$_2$S$_6^{4-}$ (5.2 nm) QDs vary in size. However, this does not account for the significant differences in $B_2$, given that the $b_2$ coefficient shows little correlation with size (Figure 3C in main text). Another distinguishing factor lies in the ligand grafting density. The CdSe-Sn$_2$S$_6^{4-}$ QDs exhibit a ligand grafting density of ~0.86 nm$^{-2}$, which is slightly larger than that of the PbS-Sn$_2$S$_6^{4-}$ QDs (~0.55 nm$^{-2}$) (see Section 6). This may result in a slightly elevated interparticle repulsion. Lastly, the Hamaker constant of CdSe (0.22 eV) is smaller than that of PbS (0.31 eV),[14, 15] implying that the QDs may experience relatively weaker Van der Waals attraction, thereby contributing to a higher $b_2$ coefficient.

### 5. DLVO theory

### 5.1 DLVO potential

The QD-QD interaction of electrostatically stabilized QDs can be modelled using Derjaguin-Landau-Verwey-Overbeek (DLVO) theory. A DLVO potential ($U_{DLVO}$) consists of repulsive screened double layer (Yukawa) potential ($U_{ES}$), attractive Van der Waals potential ($U_{VdW}$) and repulsive steric potential ($U_{steric}$).[13]

$$U_{DLVO} = U_{ES} + U_{VdW} + U_{steric}$$

The electrostatic potential can be expressed as the following, where $k_B$ is the Boltzmann constant, $T$ is the temperature, $\kappa$ is the Debye length, $R$ is the radius of the particle, $\rho_\infty$ is the bulk number density of 1:1 electrolyte, $r$ is the surface-to-surface distance between the particles.[13]

$$U_{ES}(r) = \frac{32\pi k_B T \gamma^2 \rho_\infty R}{\kappa^2} \cdot e^{-\kappa r}$$

The $\gamma$ in the above equation is described by:

$$\gamma = \tanh\left(\frac{ze\varphi}{4k_B T}\right)$$

Where $z$ is the valency of the added salt, $e$ is electron charge and $\varphi$ is the surface potential of the particle. The Debye length ($\kappa$) can be approximated with the following equation:

$$\kappa^{-1} = \sqrt{\frac{\varepsilon_0 \varepsilon_{sol} k_B T}{2e^2 I}}$$

Where $\varepsilon_0$ is the vacuum permittivity, $\varepsilon_{sol}$ is the relative dielectric constant of the solvent and $I_{sol}$ is the ionic strength of the solvent. The ionic strength is defined as the following, where $z_j$ is the charge of an ion in the solution, and $\rho_j$ is the number density of the corresponding ion.

$$I_{sol} = \frac{1}{2}\sum_{j=1}^{N} \rho_j z_j^2$$

The Van der Waals potential of two spherical particles can be described by:

$$U_{VdW} = -\frac{A}{6}\left(\frac{2R^2}{r^2 + 4rR} + \frac{2R^2}{r^2 + 4rR + 4R^2} + \ln\left(\frac{r^2 + 4rR^2}{r^2 + 4rR + 4R^2}\right)\right)$$

Where $A$ is the Hamaker constant, $R$ is the radius of the particle, $r$ is the surface-to-surface distance between the particles.[13] The steric potential was modelled using an arbitrary exponential function.

## 5.2 QD-QD potential from DLVO theory

The DLVO potentials of the QDs was calculated to gain qualitative insights into how the interparticle potential is affected by factors such as QD size, composition, surface chemistry and solvent conditions. The Hamakers constant of the QDs (PbS and CdSe)[14, 15] and the dielectric constants of the solvents (NMF, NMPA, NMF/DMF mixture)[16, 17] were sourced from available literature. The temperature used for these calculations was consistently set to $T = 298\ K$. As the DLVO theory is inherently limited to a solution of $z{:}z$ electrolyte (e.g. NaCl, CuSO4), we assume that the QDs are dispersed in a solution of $z{:}z$ electrolyte, where the ionic strength is equivalent to a solution of 0.5 mM of unbound ligand ($K_4Sn_2S_6$, $K_3AsS_4$). A concentration of 1 mM is a reasonable estimate for the concentration of unbound ligand in a ~10 mg/ml solution of PbS QDs (refer to Section 6). The value of $\gamma$ quickly approaches 1 when $\varphi$ is large, so it is assumed to be 1 for all calculations.

The DLVO potentials of PbS-$Sn_2S_6^{4-}$ QDs with diameters between 4.1 and 8.1 nm show that the strength of repulsion increases with increasing size of the QDs, which aligns with the experimental observations shown in Figure 3C of the main text.

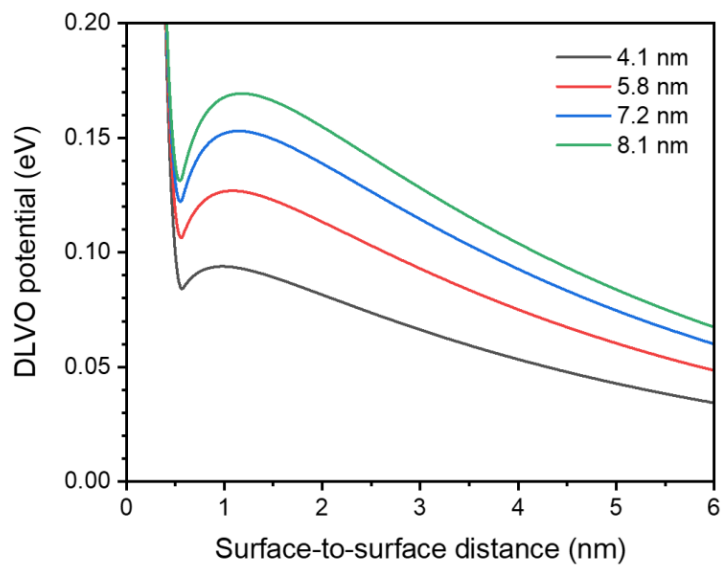

**Figure S17.** DLVO potential of PbS-$Sn_2S_6^{4-}$ QDs in NMF with the diameter ($2R$) of 4.1 nm, 5.8 nm, 7.2 nm and 8.1 nm. All solutions are assumed to contain 1 mM of $K_4Sn_2S_6$.

Following that, the comparison between PbS-$Sn_2S_6^{4-}$ and PbS-$AsS_4^{3-}$ QDs is shown below (Figure S18), where ionic strength ($I$) is changed and all other parameters are maintained constant. Due to the higher valency and ionic strength of $K_4Sn_2S_6$ compared to $K_3AsS_4$, PbS-$AsS_4^{3-}$ QDs show stronger repulsive interactions than do PbS-$Sn_2S_6^{4-}$ QDs. This agrees with the trend in the results shown in Figure 3C of the main text.

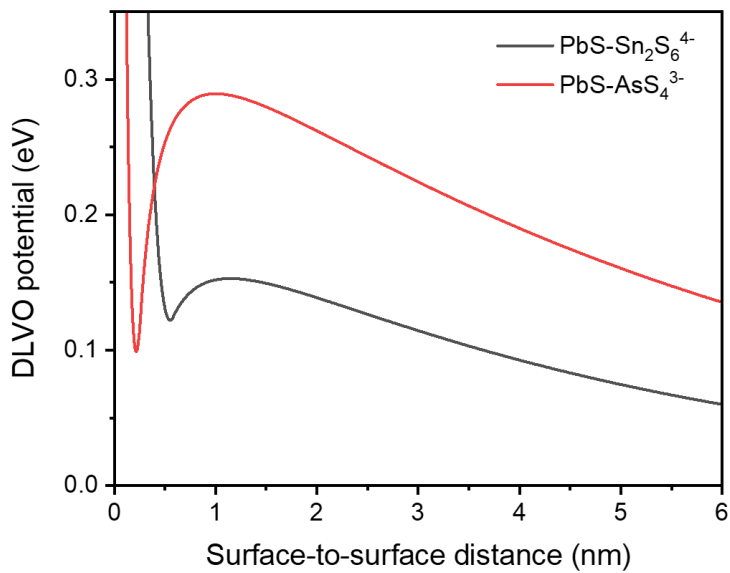

**Figure S18.** DLVO potential of 7.2 nm PbS-$Sn_2S_6^{4-}$ ($I = 6.02 \times 10^{18} \text{cm}^{-3}$) and PbS-$AsS_4^{3-}$ ($I = 3.61 \times 10^{18} \text{cm}^{-3}$) QDs in NMF. The solutions of PbS-$Sn_2S_6^{4-}$ and PbS-$AsS_4^{3-}$ QDs are assumed to contain 1 mM of $K_4Sn_2S_6$ or $K_3AsS_4$, respectively.

The DLVO potentials of PbS-$Sn_2S_6^{4-}$ QDs in three different solvent systems (NMF, NMPA, NMF/DMF mixture) indicates increased repulsion in solvents with higher dielectric constants ($\varepsilon_{sol}$), aligning with experimental observations from SAXS analysis (Figure 3C of the main text).

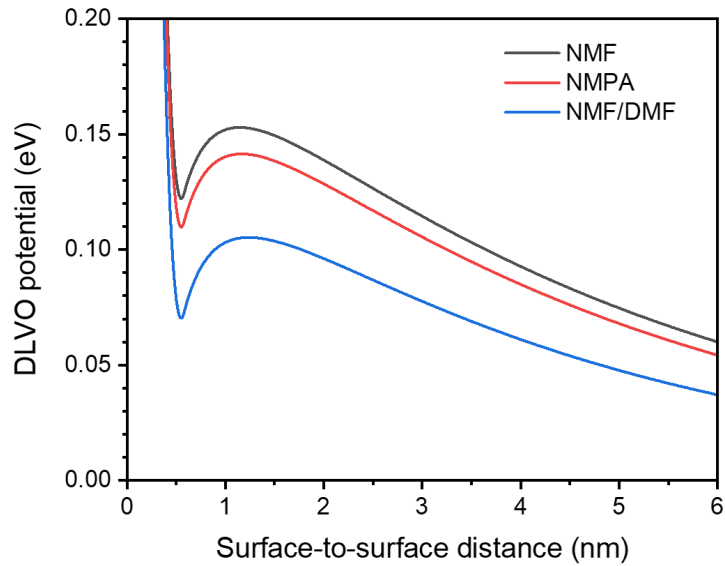

**Figure S19.** DLVO potential of 7.2 nm PbS-$Sn_2S_6^{4-}$ QDs in NMF ($\varepsilon_{sol}$ = 173), NMPA ($\varepsilon_{sol}$ = 163) and 70% NMF, 30% DMF mixture ($\varepsilon_{sol}$ = 131). All solutions are assumed to contain 1 mM of $K_4Sn_2S_6$.

Next, the DLVO potentials of PbS-$Sn_2S_6^{4-}$ and CdSe-$Sn_2S_6^{4-}$ QDs are presented below (Figure S20), with the Hamaker constant ($A$) being changed while keeping all other parameters constant. The slight variations in the Hamaker constant between the two QDs are insufficient to explain a significantly higher repulsive forces found in CdSe-$Sn_2S_6^{4-}$ QDs. The reason behind this difference remains unclear.

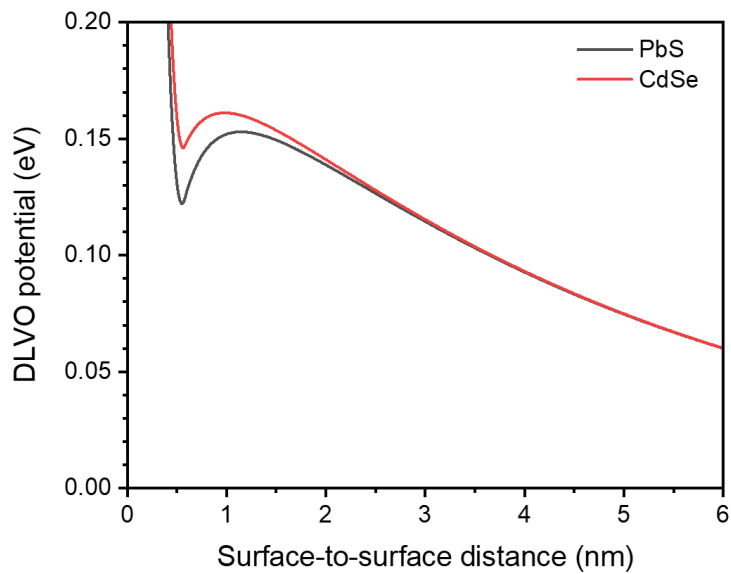

**Figure S20.** DLVO potential of 7.2 nm PbS-Sn$_2$S$_6^{4-}$ ($A = 0.311$ eV) and 7.2 nm CdSe-Sn$_2$S$_6^{4-}$ ($A = 0.219$ eV) QDs in NMF. All solutions are assumed to contain 1 mM of K$_4$Sn$_2$S$_6$.

### 5.3 Calculation of B$_2$ coefficients from DLVO theory

The $B_2$ coefficient can be calculated from the following equation, where $u(r)$ is the interparticle potential. The interparticle potential is modelled using the DLVO potential ($U_{DLVO}$).

$$B_2 = -2\pi \int_\delta^\infty \left( \exp\left(-\frac{u(r)}{k_B T}\right) - 1 \right) r^2 \, dr$$

The normalized $b_2$ coefficient can be calculated as $b_2 = B_2/B_{2,HS}$, where $B_{2,HS}$ is $B_{2,HS} = 2\pi(d + 2\sigma)^3/3$. In this equation, $d$ is the diameter of the core-only QD and $\sigma$ is the length of the ligand. $\delta$ is an arbitrary value that is close to 0. For our integral, we set it as $\delta = 0.001$ nm, which is a sufficiently small value for a reasonable estimation of $B_2$ coefficient yet large enough ensure the convergence of integral. Below is the list of $B_2$ coefficients that were obtained experimentally and from the DLVO potentials.

**Table S1.** List of experimentally and theoretically obtained $B_2$ coefficients.

| Sample | Exp. $B_2$ coefficient ($10^7$ Å$^3$) | Exp. $b_2$ coefficient | DLVO $B_2$ coefficient ($10^7$ Å$^3$) | DLVO $b_2$ coefficient |
|---|---|---|---|---|
| 7.2 nm PbS-OA, MCH | 0.27 | 0.59 | N/A | N/A |
| 4.1 nm PbS-Sn$_2$S$_6^{4-}$, NMF | 0.71 | 23.26 | 0.46 | 15.09 |
| 5.8 nm PbS-Sn$_2$S$_6^{4-}$, NMF | 2.28 | 32.31 | 0.61 | 8.56 |
| 7.2 nm PbS-Sn$_2$S$_6^{4-}$, NMF | 2.64 | 21.58 | 0.71 | 5.83 |
| 8.1 nm PbS-Sn$_2$S$_6^{4-}$, NMF | 4.93 | 27.80 | 0.78 | 4.68 |
| 7.2 nm PbS-Sn$_2$S$_6^{4-}$, NMF/DMF | 1.88 | 15.40 | 0.38 | 3.10 |
| 7.2 nm PbS-Sn$_2$S$_6^{4-}$, NMPA | 2.03 | 16.60 | 0.62 | 5.10 |
| 7.2 nm PbS-AsS$_4^{3-}$, NMF | 6.84 | 70.96 | 2.22 | 23.05 |
| 5.6 nm CdSe-Sn$_2$S$_6^4$, NMF | 6.28 | 116.59 | 0.71 | 5.84 |

## 6. Analysis of grafting density of Sn$_2$S$_6^{4-}$ and AsS$_4^{3-}$ ligands

The grafting densities of MCC ligands (Sn$_2$S$_6^{4-}$ and AsS$_4^{3-}$) on PbS and CdSe QDs were determined by elemental analysis.

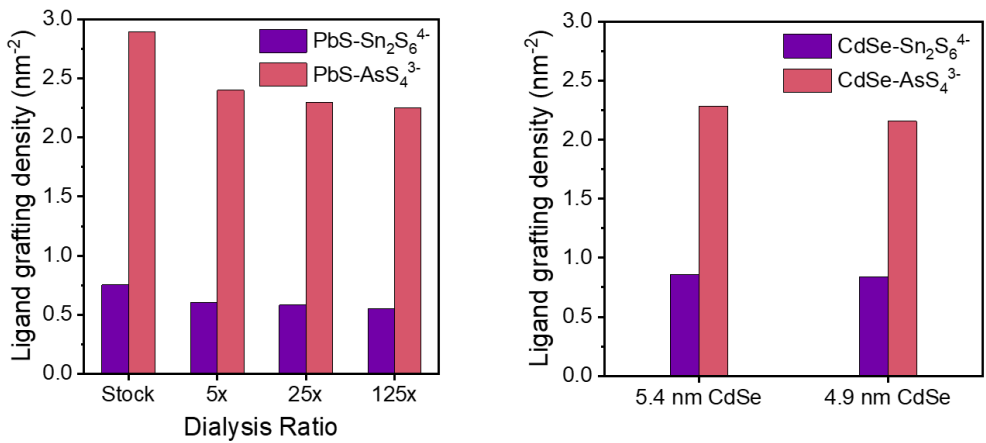

**Figure S21.** Grafting densities of $Sn_2S_6^{4-}$ and $AsS_4^{3-}$ ligands on 5.8 nm PbS QDs following a series of dialysis cycles (left); grafting densities of $Sn_2S_6^{4-}$ and $AsS_4^{3-}$ ligands on CdSe QDs with two different sizes (right).

Figure S21 illustrates that after the x125 dialysis, the number of $Sn_2S_6^{4-}$ ligands on 5.8 nm PbS QDs decreased from 0.75 to 0.55 per nm$^2$ and the number of $AsS_4^{3-}$ ligands decreased from 2.89 to 2.25 per nm$^2$. This indicates that the remaining number of $Sn_2S_6^{4-}$ and $AsS_4^{3-}$ ligands on PbS QDs is approximately 0.55 nm$^{-2}$ and 2.25 nm$^{-2}$, respectively, after most free ligands were removed from the solution. However, it is essential to note that the loss of ligands from the QDs is not solely due to initially free ligands, as the equilibrium distribution between free and bound ligands can change, leading to partial desorption of bound ligands from the QD surface. Consequently, the calculated ligand grafting density after the series of dialysis may be overestimated.

The grafting density of $Sn_2S_6^{4-}$ ligands on 5.2 nm CdSe was found to be 0.86 nm$^{-2}$, which is reasonably consistent with a previous report.[18] Elemental analysis of two sizes of CdSe QDs in NMF shows a similar ligand grafting density for both $Sn_2S_6^{4-}$ and $AsS_4^{3-}$ ligands.

## 7. Structure analysis of the films

### 7.1 Bragg peak analysis of PbS-OA and PbS-$Sn_2S_6^{4-}$ superlattices

The scattering vectors of the Bragg peaks ($q_{max}$) and the Miller indices ($h\ k\ l$) follow the following relationships, where $a$ is the length of a cubic unit cell.

$$q_{max} = \frac{2\pi}{a}\sqrt{h^2 + k^2 + l^2}$$

Figure S22 (right) shows the plots of the position of the $q_{max}$ and the value of $2\pi(h^2 + k^2 + l^2)^{1/2}$ of the Miller indices that best match the $q_{max}$. The length of the unit cell was obtained from the slope. Figure S22 (left) are the assignment of the peaks. Note that some peaks are not visible due to the suppression of the peak intensity from the form factor component. From the SAXS analysis, we find that the superlattice of 7.2 nm PbS-OA QDs adopts the fcc structure with the unit cell length of 13.2 nm, 7.2 nm PbS-Sn$_2$S$_6^{4-}$ QDs adopt the fcc structure with the unit cell length of 11.0 nm, and the 7.2 nm PbS-AsS$_4^{3-}$ QDs adopt the bcc structure with the unit cell length of 8.6 nm.

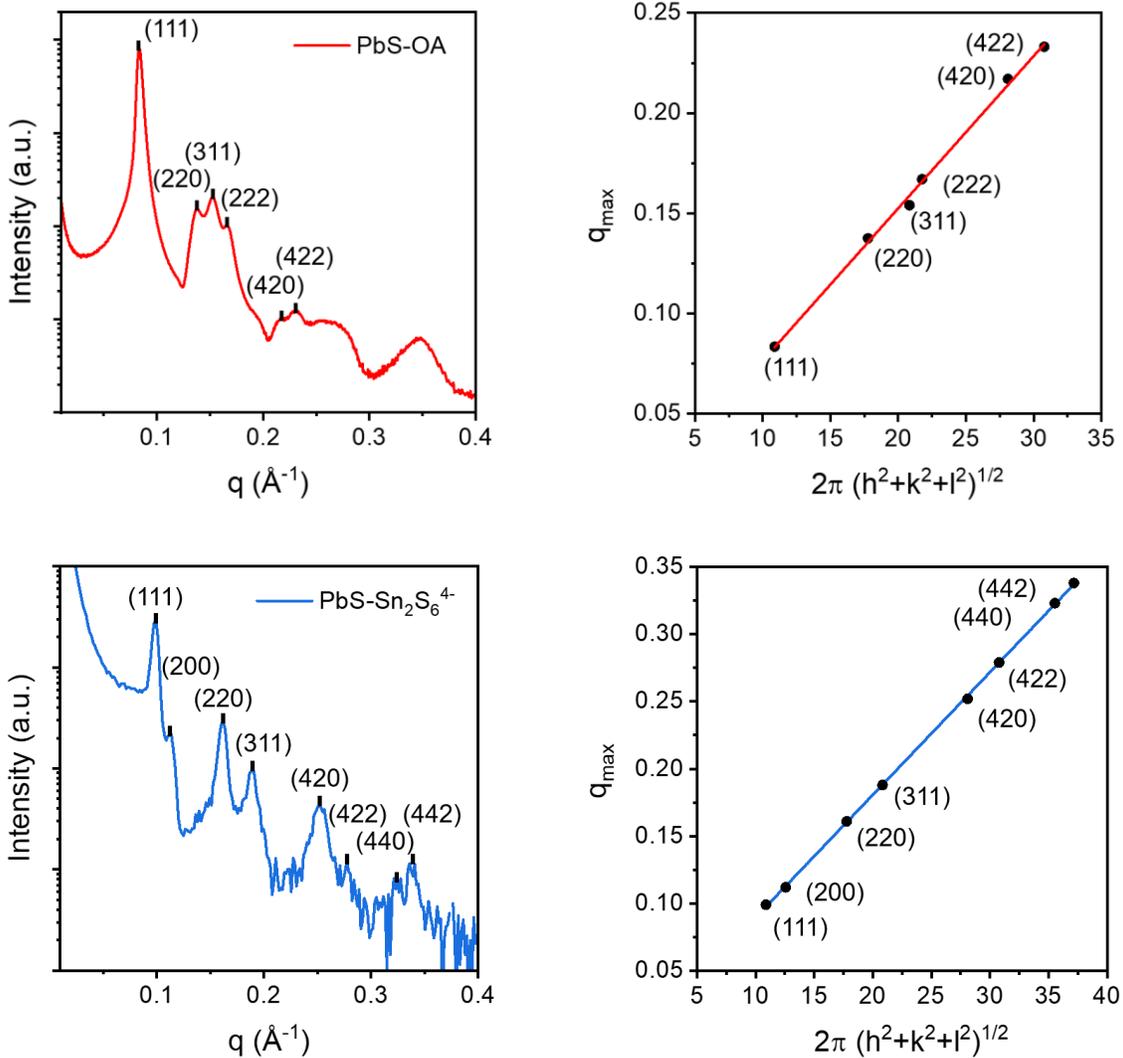

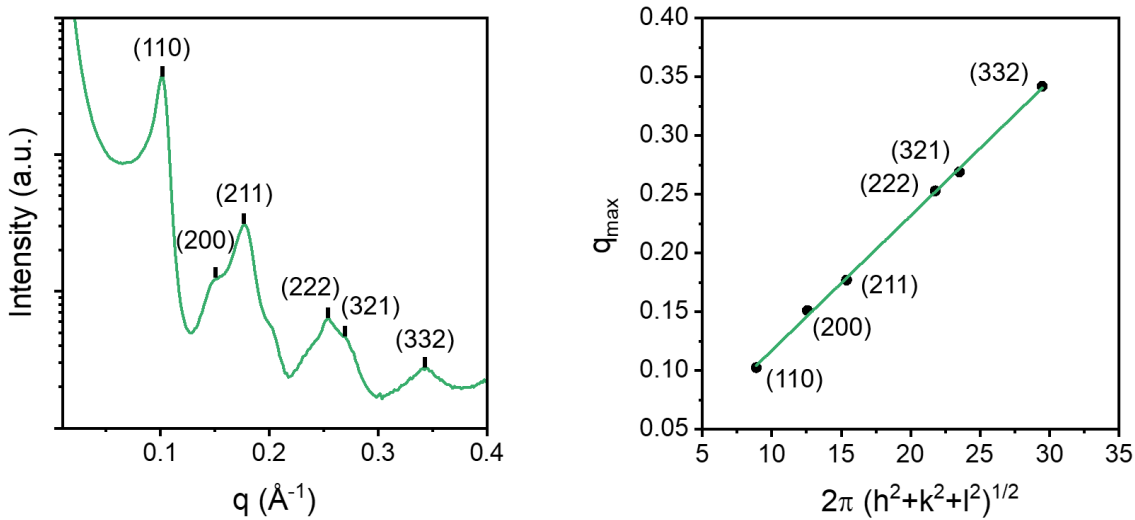

**Figure S22.** The log-lin plot (left) of SAXS pattern and the plot of $q_{max}$ against $2\pi(h^2 + k^2 + l^2)^{1/2}$ (right) of the superlattice of 7.2 nm PbS-OA (red), 7.2 nm PbS-$Sn_2S_6^{4-}$ (blue) and 7.2 nm PbS-$AsS_4^{3-}$ (green) QDs.

### 7.2 Films of spin-coated PbS QDs

Figure S23 presents the SEM images of spin-coated films of PbS QDs. SAXS patterns depicted in Figure S24 illustrate that thin films formed by spin-coating PbS-OA QDs adopt a bcc structure, while those formed by PbS-$Sn_2S_6^{4-}$ QDs exhibit an amorphous structure. When $K_3AsS_4$ is added to PbS-$Sn_2S_6^{4-}$ QDs, the thin films exhibit an fcc structure. These findings mirror those observed in drop-casted films. The lower signal-to-noise ratio observed can be attributed to the reduced thickness of the spin-coated film compared to drop-casted films.

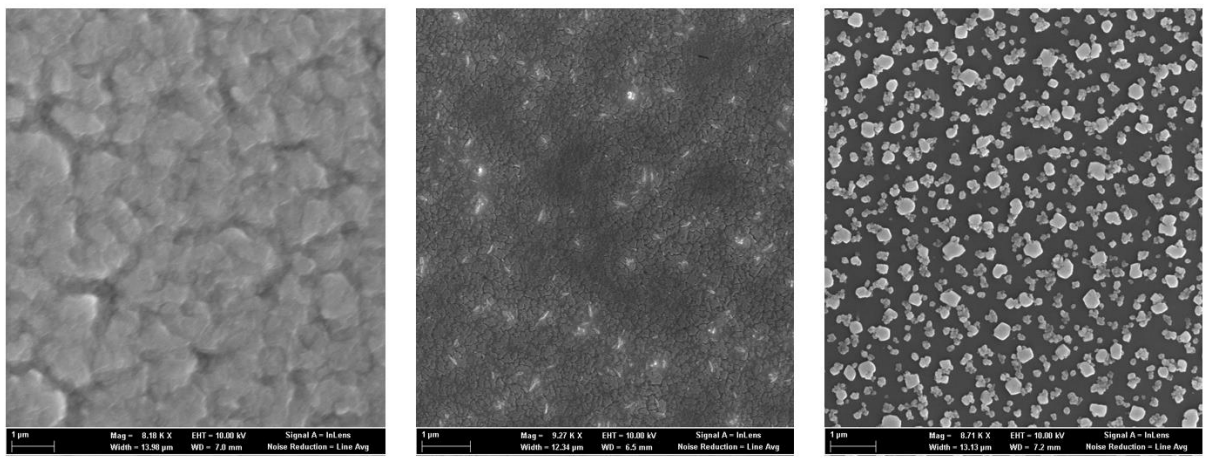

**Figure S23.** SEM image of film of PbS-OA QDs (left), PbS-$Sn_2S_6^{4-}$ QDs (center), PbS-$Sn_2S_6^{4-}$ QDs with 20 mM $K_3AsS_4$ (right). All films were prepared by spin-coating.

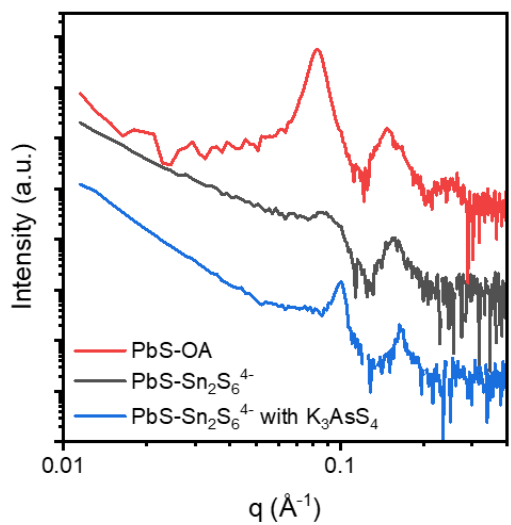

**Figure S24.** SAXS patterns of PbS-$Sn_2S_6^{4-}$ QDs spin-coated on a thin Si safer, with a varying concentration of $K_3AsS_4$ added.

### 7.3 Morphology of films

The morphology of the film of PbS-$Sn_2S_6^{4-}$ QDs can be seen from the SEM images below. This film was prepared by drop-casting a solution of PbS-$Sn_2S_6^{4-}$ QDs in NMF on a Si wafer.

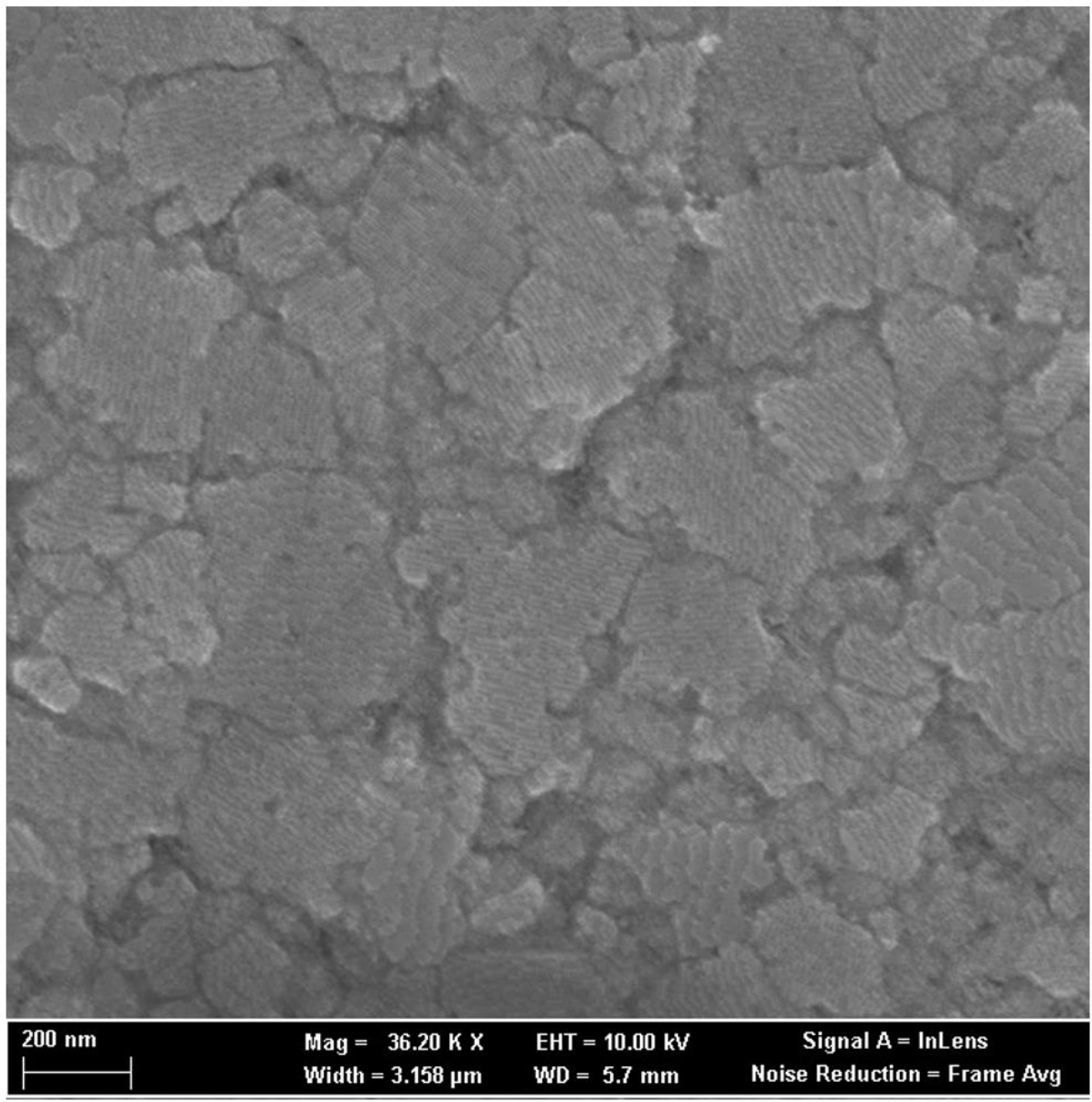

**Figure S25.** SEM image of film of PbS-$Sn_2S_6^{4-}$ QDs.